\documentclass[aps,epsfig,showkeys,letter,nofootinbib,superscriptaddress,eqsecnum,]{revtex4} %
\usepackage{pstricks}
\usepackage[dvips]{graphicx} 
\usepackage{amssymb}
\usepackage{bm}
\usepackage{rotate}

\newcommand{\be}{\begin{equation}}
\newcommand{\ee}{\end{equation}}
\newcommand{\bea}{\begin{eqnarray}}
\newcommand{\eea}{\end{eqnarray}}
       
\newcommand{\refeq}[1]{Eq.~(\ref{eq:#1})}          
\newcommand{\refeqs}[2]{Eqs.~(\ref{eq:#1})--(\ref{eq:#2})}          
          
\newcommand{\reffig}[1]{Fig.~\ref{fig:#1}}          
\newcommand{\reftab}[1]{Tab.~\ref{tab:#1}}

\newcommand{\refsec}[1]{Section~\ref{sec:#1}}          
\newcommand{\refapp}[1]{Appendix~\ref{app:#1}}

\renewcommand{\v}[1]{\mathbf{#1}}

\newcommand{\ph}{\varphi}

\newcommand{\eps}{\varepsilon}

\newcommand{\f}[1]{{#1}^{(4)}}
\newcommand{\ff}[1]{{#1}^{(5)}}

\newcommand{\Mplf}{M_{\rm Pl\,(4)}}
\newcommand{\Mplff}{M_{\rm Pl\,(5)}}
\newcommand{\nablat}{\tilde{\nabla}}

\newcommand{\Mpch}{\,{\rm Mpc}/h}
\newcommand{\iMpch}{\,h/{\rm Mpc}}
\newcommand{\Lbox}{L_{\rm box}}
\newcommand{\kNy}{k_{\rm Ny}}
\newcommand{\kmax}{k_{\rm max}}
\newcommand{\Msunh}{\,M_{\odot}/h}

\newcommand{\Om}{\Omega_m}
\newcommand{\Ob}{\Omega_b}
\newcommand{\OL}{\Omega_\Lambda}
\newcommand{\Orc}{\Omega_{\rm rc}}
\renewcommand{\L}{\mathcal{L}}

\begin{document}


\title{Self-Consistent Cosmological Simulations of DGP Braneworld Gravity}

\author{Fabian Schmidt}
\affiliation{Department of Astronomy \& Astrophysics, The University of
Chicago, Chicago, IL 60637-1433}
\affiliation{Kavli Institute for Cosmological Physics, Chicago, IL 
60637-1433}

\begin{abstract}
We perform cosmological N-body simulations of the Dvali-Gabadadze-Porrati
braneworld model, by solving the full non-linear equations of motion
for the scalar degree of freedom in this model, the brane bending mode.
While coupling universally to matter, the brane-bending mode has
self-interactions that become important as soon as the density field
becomes non-linear. These self-interactions lead to a suppression
of the field in high-density environments, and restore gravity
to General Relativity. The code uses a multi-grid relaxation scheme 
to solve the non-linear field equation in the quasi-static approximation.
We perform simulations of a flat self-accelerating DGP model without 
cosmological constant.
However, the type of non-linear interactions of the brane-bending mode,
which are the focus of this study, are generic to a wide class of 
braneworld cosmologies.
The results of the DGP simulations are compared with standard gravity
simulations assuming the same expansion history, and with
DGP simulations using the linearized equation for the brane bending
mode. This allows us to isolate the effects of the non-linear self-couplings
of the field which are noticeable already on quasi-linear scales.
We present results on the matter power spectrum and the halo mass function,
and discuss the behavior of the brane bending mode within cosmological
structure formation.
We find that, independently of CMB constraints, the self-accelerating DGP 
model is strongly
constrained by current weak lensing and cluster abundance measurements.
\end{abstract}


\date{\today}

\maketitle

\section{Introduction}

The observed acceleration of the universe 
\cite{RiessEtal,KowalskiEtal,WMAP5,Giannantonio,Pietrobon}
poses a fascinating
challenge to physicists and cosmologists: it points towards
new physics at an unusually low energy scale ($\sim 10^{-3}\,$eV),
or large length scale ($\sim 10^3\,$Mpc). For this reason,
no natural explanation appears to exist up to now. Many attempts
have been made to go beyond the minimal explanation,
a cosmological constant or vacuum energy. They can be broadly 
classified into two categories: 
acceleration is due to an additional,
smooth stress-energy component with negative pressure (\textit{dark
energy} \cite{FriemanReview}); or it is caused by gravity itself, through 
modifications to 
General Relativity (GR) on cosmological scales.
While there are strong constraints on deviations from GR in the
Solar System \cite{Will}, gravity is remarkably weakly constrained
on cosmological length scales. This provides an independent motivation
for studying modified gravity models in the context of cosmology.

Smooth dark energy models have enough freedom to reproduce essentially
any background expansion history of the universe. 
In order to distinguish modified gravity
from the smooth dark energy scenario, it is thus necessary to study
the growth of cosmological structure. A wealth of observables
can be used for this purpose
\cite{ZhangEtal,JainZhang,SongKoyama,KnoxSongTyson,Schmidt08,Song2005,Song2006,JainZhang,Tsujikawa2008,ZhangfR,SongEtalDGP,Schmidt07}, e.g. weak lensing, 
galaxy-CMB cross-correlation,
galaxy cluster abundances, and many more. However, almost all of these
measurements are affected by non-linearities in the matter density field,
or have most of their information on non-linear scales. All viable
modified gravity models include a non-linear mechanism to restore
General Relativity in high-density environments, which is necessary
in order to satisfy Solar System constraints. This mechanism has to
be taken into account in order to make accurate
predictions for observables on non-linear scales in modified gravity. 
For one representative of $f(R)$ modified gravity, 
\cite{HPMpaperII,HPMhalopaper} showed that the effects of this non-linear
{\it chameleon} mechanism are significant for models that satisfy
Solar System constraints.

One of the most popular modified gravity scenarios is the 
Dvali-Gabadadze-Porrati (DGP) model \cite{DGP1}. In this model, matter
and radiation are confined to a 4-dimensional brane in a 5-dimensional
bulk spacetime. While gravity propagates in five dimensions on the
largest scales, it becomes 4-dimensional below a certain 
\textit{crossover scale} $r_c$. 
On scales smaller than $r_c$, and when gravity is weak, DGP gravity can be 
described as an effective four-dimensional scalar-tensor theory \cite{NicRat}. 
The scalar field, referred to as \textit{brane bending mode}, is 
massless but has non-linear derivative interactions
which suppress the field in high-density environments, restoring GR locally
and allowing DGP gravity to pass Solar System constraints.

Applied to homogeneous and isotropic cosmology, the DGP model allows for
two solutions \cite{Deffayet01}. In one branch of the theory, the 
\textit{self-accelerating branch}, the effective weakening of gravity on large 
scales leads to an accelerated late-time expansion of the brane,
without any cosmological constant. The scalar brane-bending mode mediates
a repulsive force, leading to a smaller effective gravitational constant
on large scales. In the other solution, the \textit{normal branch}, there
is no accleration, and the brane-bending mode mediates an attractive force.
In linear perturbation theory around its de~Sitter limit,
the brane bending mode in the self-accelerating DGP model has been shown to 
have the wrong sign in the kinetic term (``ghost'')
\cite{LutyEtal,NicRat}, suggesting an instability, although the situation in 
the non-linear case is not clear \cite{Dvali06,KoyamaReview}. The normal 
branch is free of the ghost.

Linear perturbation theory around a cosmological background in DGP
has been derived in, e.g. \cite{KoyamaMaartens,SongEtalDGP,CardosoEtal}.
In this approach, it is possible to predict
the expansion history, CMB anisotropies, and the ISW effect 
in DGP \cite{FangEtal1}. 
Using this information, \cite{FangEtal} have shown
that the self-accelerating DGP model without cosmological constant,
which has only one free parameter, $r_c$,
is disfavored at the $\sim 4\sigma$ level by current CMB and Supernova
data. This is due to the earlier onset of acceleration in DGP, and the 
additional
suppression of growth through the brane-bending mode.

While the self-accelerating branch of DGP is thus challenged by theory
and observations, much work is being done on extending the DGP model to
higher co-dimensions, in the context of degravitation \cite{deGrav,deRham}.
These higher-dimensional models are expected to bring the expansion
history close to that of $\Lambda$CDM, while exhibiting a similar
effective scalar-tensor regime as in the original DGP model (see also
\cite{AfshordiEtal}).
The form of the non-linear interactions of the brane-bending mode are
also generic to these generalized models (see also \cite{galileon}). 
Hence, it will be straightforward
to perform simulations of these models once the expansion history and
evolution of linear perturbations is worked out.
In addition, the standard DGP model is able to satisfy the cosmological 
constraints if a cosmological constant (brane tension) is included
\cite{LombriserEtal}.

So far, studies of the DGP model in the context of cosmology have 
mostly dealt with 
the linearized theory, valid on large scales. However, as soon as 
perturbations in the matter density become of order unity, the non-linear
interactions of the brane bending mode become important 
(e.g., \cite{KoyamaSilva}). Hence, in order to make predictions for observables
in the non-linear regime, the full non-linear equations of motion must
be solved in conjunction with the evolution of the matter
perturbations. Recently, perturbative approaches have been developed
to extend the predictions into the quasi-linear regime \cite{KoyamaEtal09}.
Also, Khoury and Wyman \cite{KW} have used a spherically-symmetric
approximation of the brane-bending mode interactions in their
N-body simulation of braneworld models.
In this paper, we present the results of an N-body simulation
of the self-accelerating DGP model,
which self-consistently solves the {\it full} non-linear equation for the
brane-bending mode and its effect on the motion of particles
(we compare our results with the approximation used in \cite{KW} in 
\refsec{profiles}).
We solve the equation of motion in the quasi-static approximation,
neglecting time derivatives with respect to spatial derivatives,
as is usually done in N-body simulations (\refsec{qstest} presents
a consistency test of this approximation).

We present results on the matter power spectrum, the halo mass function
(abundance of dark matter halos), and the behavior of the brane
bending mode in the cosmological context. These results can be used
to strengthen constraints on the self-accelerating DGP model significantly,
e.g., through observations of weak lensing and the abundance of galaxy 
clusters. More generally, the simulations presented here can serve as
starting point for building a model of non-linear structure formation
in braneworld scenarios, and for benchmarking perturbative approaches
\cite{KoyamaEtal09}.

In \refsec{DGP}, we describe the DGP model, and present the relevant
equations and analytical test cases. 
\refsec{code} presents the code, which is benchmarked
in \refsec{tests}. The cosmological simulations are described in
\refsec{sims}. Results and their impact on cosmological constraints
on the DGP model are presented in \refsec{res} and \refsec{constraints}.
We conclude in \refsec{concl}.

\section{DGP model}
\label{sec:DGP}

\subsection{Background evolution}

The Dvali-Gabadadze-Porrati model \cite{DGP1} consists of a spatially
three-dimensional
brane in a 4+1 dimensional Minkowski bulk. Matter and all interactions except
gravity are confined to the brane. The gravitational action consists of
the five-dimensional Einstein-Hilbert action plus a term which leads to 
the 4D gravity limit on small scales:
\be
S_{\rm grav} = -\frac{1}{16\pi\,\ff{G}} \int d^5 X \sqrt{-\ff{g}} \ff{R} + S_{\rm boundary}
- \frac{1}{16\pi\,\f{G}} \int d^4 x \sqrt{-\f{g}} \left ( \f{R} + \mathcal{L}_m \right ).
\label{eq:action}
\ee
Here, $X$, $\ff{g}$ stand for the bulk coordinates and metric, while $x$, $\f{g}$ are the 
induced coordinates and metric on the brane, and $\ff{R}$, $\f{R}$ denote the
corresponding Ricci scalars. In the following, we will drop the $\f{}$ notation
where no confusion can arise. The boundary term is added to the action
in order to ensure that variation with respect to $\ff{g}$ leads to
the correct five-dimensional Einstein equations (e.g., \cite{GibbonsHawking}).

The two gravitational constants, or Planck masses $\Mplff$, $\Mplf$ appearing
in \refeq{action} can be related via a length scale, the 
\textit{crossover scale}:
\be
r_c \equiv \frac{1}{2} \frac{\ff{G}}{\f{G}} = \frac{1}{2} \frac{\Mplf^2}{\Mplff^3}.
\label{eq:rc}
\ee
On scales above $r_c$, gravity becomes five-dimensional, with forces falling
off as $1/r^3$. Below $r_c$, gravity is four-dimensional, but not Einsteinian
gravity, a point to which we return below.

Since all matter is thought as confined to the brane, the five-dimensional
metric has to obey non-trivial junction conditions over the brane \cite{Deffayet01}.
Assuming an empty Minkowski bulk, a spatially flat brane,
and a homogeneous and isotropic matter
distribution on the brane, the junction conditions lead to the following
analogue of the Friedmann equation for the scale factor of the induced
metric on the brane:
\be
H^2 \pm \frac{H}{r_c} = \frac{8\pi\,G}{3} \rho, \quad H \equiv \frac{\dot a}{a}.
\label{eq:Friedmann}
\ee
The junction conditions leave two possible branches of the theory, determined
by the sign on the left hand side of \refeq{Friedmann}. In the following,
we focus on the branch with the `$-$' sign, which asymptotes to a late-time 
de Sitter-Universe,
$H=1/r_c=\rm const.$, and is correspondingly called the \textit{accelerating
branch}. 
Then, in the matter-dominated epoch, neglecting radiation, and again assuming no 
cosmological constant or curvature, 
\refeq{Friedmann} can be rewritten as:
\bea
E(a) \equiv \frac{H(a)}{H_0} &=& \sqrt{\Orc} + \sqrt{\Om a^{-3} + \Orc}, \;\mbox{where} \label{eq:DGPexp}\\
\Orc &\equiv& \frac{1}{4H_0^2r_c^2};\quad \Om \equiv \frac{8\pi G}{3 H_0^2} \rho_{m0},\nonumber
\eea
and $\rho_{m0}$ is the average matter density today. This expansion history
is clearly different from $\Lambda$CDM, and corresponds to an effective
dark energy with $w_{\rm eff} \rightarrow -1/2$ in the matter-dominated era
at high redshifts.

\subsection{Cosmological perturbations and brane bending mode}

The propagation of light and particles on the DGP brane is completely
determined by the perturbed 4D Friedmann-Robertson-Walker metric:
\be
ds^2 = -[1 + 2\Psi(\v{x},t)] dt^2 + [1 + 2\Phi(\v{x},t)] a^2(t) d\v{x}^2
\label{eq:metric}
\ee
However, in order to determine the evolution of the metric potentials
on the brane, it is necessary to solve the full 5D Einstein equations
\cite{KoyamaSilva,SongEtalDGP}. An additional scalar degree of freedom
associated with local displacements of the brane appears, the so-called
\textit{brane-bending mode} $\ph$ which couples to matter.
In our convention, $\ph$ is dimensionless, instead of being scaled
to the Planck mass $\Mplf$.

In the decoupling limit of DGP \cite{NicRat}, when setting gravitational
interactions to 0, the self-interactions of the $\ph$ field remain constant.
Hence, while perturbations of the metric higher than linear order can be
neglected for cosmological studies, 
it is crucial to consider the self-interactions in the brane-bending
mode.

In the quasi-static regime, for scales $k\gg H_0,\:r_c^{-1}$, time derivatives
can be neglected with respect to spatial derivatives, and the 
equation for the brane-bending mode reads (e.g., \cite{KoyamaSilva}):
\be
\nabla^2 \ph + \frac{r_c^2}{3\beta\,a^2} [ (\nabla^2\ph)^2
- (\nabla_i\nabla_j\ph)(\nabla^i\nabla^j\ph) ] = \frac{8\pi\,G\,a^2}{3\beta} \delta\rho,
\label{eq:phiQS}
\ee
where the derivatives are with respect to comoving coordinates $\v{x}$,
$\delta\rho = \rho_m-\overline{\rho_m}$ is the matter density perturbation,
and the function $\beta(a)$ (for the accelerating branch) is given by:
\be
\beta(a) = 1 - 2 H(a)\, r_c \left ( 1 + \frac{\dot H(a)}{3 H^2(a)} \right ).
\label{eq:beta}
\ee
Note that $\beta$ is always negative in the self-accelerating branch.
In \refsec{qstest}, we show results of a consistency test of the 
quasi-static assumption used in \refeq{phiQS}.

The brane-bending mode influences the dynamics of particles through
the dynamical potential $\Psi$, which, assuming the same boundary
conditions for $\Psi$ and $\ph$, is given by:
\be
\Psi = \Psi_N + \frac{1}{2} \ph,
\label{eq:psi}
\ee
where the Newtonian potential $\Psi_N$ satisfies the usual Poisson equation:
\be
\nabla^2\Psi_N = 4\pi G\,a^2\,\delta\rho.
\ee
In contrast, the propagation of photons, determined by the lensing
potential $\Phi_- \equiv (\Phi-\Psi)/2$, is not directly affected by $\ph$.

\subsection{Weak brane regime}
\label{sec:philin}

If the gradient of the $\ph$ field is small, i.e. for small gravitational
accelerations, \refeq{phiQS} can be linearized, yielding a standard
Poisson equation:
\be
\quad \nabla^2\ph_{L} = \frac{8\pi G\,a^2}{3\beta} \delta\rho
\quad\Rightarrow\quad \ph_L = \frac{2}{3\beta}\Psi_N.
\label{eq:philin}
\ee
We will refer to a cosmology with this linearized equation as the 
\textit{linearized DGP model}.
In this regime, also called weak-brane phase, 
$\ph=\ph_L$ becomes proportional to the Newtonian potential
$\Psi_N$, corresponding to a constant rescaling of the gravitational
constant through \refeq{psi}:
\be
G_N \rightarrow G_{\rm eff} = \left ( 1 + \frac{1}{3\beta} \right) G_N.
\label{eq:Geff}
\ee
Substituting the linear solution into \refeq{phiQS}, we obtain
a rough estimate for the overdensity $\delta\equiv\delta\rho/\bar\rho_m$
at which the non-linear interactions become important:
\be
1 \sim\frac{r_c^2}{3\beta\,a^2}(\nabla^2\ph_L)^2 \left / \right . \nabla^2\ph_L 
= \frac{8\pi G\,r_c^2}{9\beta^2}\delta\rho
= \frac{3 H_0^2 r_c^2}{9\beta^2}\Omega_m\,\delta.
\label{eq:NLest}
\ee
For a self-accelerating DGP model that leads to cosmic acceleration today,
$r_c\sim H_0^{-1}$, $\beta(a=1)\sim -1$, and the prefactor in \refeq{NLest} is 
of order unity today.
Hence, the self-coupling of the brane bending mode $\ph$ becomes important
as soon as the matter density field becomes non-linear, $\delta\sim 1$.

\subsection{Vainshtein effect}
\label{sec:vainshtein}

In general, \refeq{phiQS} is difficult to solve in full generality 
due to the non-linearity in the derivative terms. However, an instructive 
test case, a spherically 
symmetric matter distribution, can be solved analytically.
In the spherically symmetric case, \refeq{phiQS} becomes \cite{KoyamaSilva}:
\bea
\left ( \frac{d^2}{dr^2} + \frac{2}{r}\frac{d}{dr} \right )
(\ph + \frac{1}{3\beta\, a^2} \Xi ) &=& \frac{8\pi G\,a^2}{3\beta} \delta\rho,
\label{eq:phiSph} \\
\Xi \equiv 2 r_c^2 \int_0^r \frac{dr'}{r'} \left (\frac{d\ph}{dr'} \right )^2. & & \nonumber
\eea
For simplicity, we assume a spherical mass $M$ of radius $R$ with
uniform density, and set $a=1$. 
Then, we can integrate \refeq{phiSph} once and obtain
the gravitational acceleration in DGP:
\be
g = g_N + \frac{1}{2}\frac{d\ph}{dr} = g_N [1 + \Delta(r)],
\label{eq:g_sph}
\ee
where $g_N$ is the Newtonian acceleration of the spherical mass, and:
\be
\Delta(r) =  \frac{2}{3\beta} \left \{ 
\begin{array}{rl}
r^3/r_*^3 \left ( \sqrt{1 + \left ( \frac{r_*}{r} \right )^3} - 1 \right ), & r \geq R \vspace*{0.2cm}\\
R^3/r_*^3 \left ( \sqrt{1 + \left ( \frac{r_*}{R} \right )^3} - 1 \right ), & r < R.
       \end{array} \right .
\label{eq:Delta}
\ee
Here, $r_*$ denotes a characteristic scale of the solution, the 
\textit{Vainshtein radius}:
\be
r_*^3 = \frac{8\,r_c^2 r_s}{9\beta^2},
\label{eq:rstar}
\ee
and $r_s = 2G M $ is the gravitational radius of the mass. For very large
distances, $r \gg r_*$, $\Delta(r)$ approaches the constant $1/(3\beta)$,
which exactly matches the linear solution, \refeq{Geff}. Note also that
by substituting $\delta\rho\sim M/r^3$ in \refeq{NLest}, we see that
the non-linearity criterion is directly proportional to $(r_*/r)^3$.

In the opposite limit, $r \ll r_*$, $\ph$ becomes suppressed with 
respect to the linear solution, and $\Delta(r)$ approaches the small constant
$2/(3\beta) (R/r_*)^{3/2}$ inside the mass (assuming $R \ll r_*$).
This \textit{Vainshtein effect} \cite{Vainshtein72,DeffayetVainshtein02} 
amounts to restoring GR in deep potential 
wells, while far away from the mass, gravity is in the scalar-tensor regime.

\reffig{sphericalmass} shows the analytical solution for the $\ph$ field
and the deviation from the Newtonian acceleration, as well as the code results
which agree well with the analytical solution in the regime of validity
(see \refsec{sphericalmass}).

\subsection{Plane wave solution}
\label{sec:planewave}

Another exact solution to \refeq{phiQS} exists: for a simple plane wave
density perturbation, $\delta\rho(\v{x}) = A\,\exp(i \v{k}\cdot\v{x})$.
In this case, the two non-linear terms exactly cancel, and we are left
with the linear solution, \refeq{philin}.
In \refapp{sinetest} this is used to test our code for perturbations
of different wavelengths.
Also, the plane wave and spherically symmetric solutions can be considered 
as two limiting cases for understanding the $\ph$ field behavior in the
cosmological context.
We will return to this point in \refsec{profiles}.

\section{Code implementation}
\label{sec:code}

We use an extension of the code first described in \cite{HPMpaper} for
our simulations of DGP gravity. The code is a standard 
particle-mesh N-body code \cite{HockneyEastwood,KlypinHoltzman} 
assuming collisionless dark matter only, with a
fixed grid of size $N_g^3$ and periodic boundary conditions, and uses 
second-order accurate leapfrog integration for the 
particle propagation.
The major addition is a relaxation solver for the non-linear
equation of the brane bending mode $\ph$, \refeq{phiQS}. The simulation
proceeds by performing steps $\Delta a$ in the scale factor $a$. 

Except for the different expansion history, the equations for particle
propagation are identical to those in standard CDM simulations, and
are expressed in terms of comoving coordinates $\v{x}$, $\v{p}$:
\bea
\frac{d\v{x}_i}{da} &=& \frac{1}{a^3\,E(a)}\v{p}_i, \\
\frac{d\v{p}_i}{da} &=& -\frac{1}{a\,E(a)} \bm{\nabla} \Psi(\v{x}_i).
\eea
Here, $\v{x}_i$ and $\v{p}_i$ are the position and momentum of particle
$i$ in code units (see \refapp{code}), and $a\,E(a)$ is taken from 
\refeq{DGPexp}. The main modification enters in the 
determination of $\Psi(\v{x})$.
One time step of the N-body simulation proceeds as follows:
\begin{itemize}
\item The density field on a fixed grid is assigned using the particle positions at scale
factor $a$ via the cloud-in-cell method, i.e. each particle corresponds to 
a uniform density cube of side length $r_{\rm cell}$ and mass $M = \bar\rho_m\,r_{\rm cell}^3$.
\item The Newtonian potential $\Psi_N$ is solved for using a Fast Fourier
Transform; \refeq{phiQS} is solved via relaxation, and the overall dynamical
potential $\Psi$ is assembled according to \refeq{psi}. We also run
linearized DGP simulations, where the potential is simply given by
$\Psi = [1 + 1/(3\beta(a))]\,\Psi_N$.
\item Using the same cloud-in-cell interpolation as for particles, the
gradient of $\Psi$ is calculated at each particle position and is used to
update the particle momenta from $a-\Delta a/2$ to $a+\Delta a/2$.
\item Using the momenta at $a+\Delta a/2$, the particle positions are
updated from $a$ to $a + \Delta a$, and the process starts from the
beginning.
\end{itemize}

Thus, in each time step we have to solve for
the brane-bending mode $\ph$ which contributes to the dynamical potential
$\Psi$ via \refeq{psi}. In other words, given the density field, we have to solve
the non-linear elliptical differential equation \refeq{phiQS}. We
use a Gauss-Seidel relaxation scheme together with the Newton method. A 
crucial tool is the {\it multigrid}
\cite{Brandt2,Brandt3,Briggs},
a hierarchy of coarser grids which are essential to speeding up the 
convergence \cite{HPMpaper}: as the relaxation scheme, which operates locally,
gets rid of small-scale errors very efficiently, the coarser grids
quickly reduce the long-wavelength error modes which are hard to eliminate
on the fine grid alone. For details of the implementation, see
\refapp{code}.

While the $\ph$ field is in general well-behaved, the convergence properties
become worse for strongly inhomogeneous density fields, because the 
non-linearities in \refeq{phiQS} are in the {\it derivatives} of the $\ph$
field. In order to reach the desired convergence in our cosmological 
simulations, it is necessary to smooth the density field entering the
r.h.s. of \refeq{phiQS} with a Gaussian kernel $\propto \exp (-r^2/2r_s^2)$,
with $r_s$ set to the size of a grid cell.
This smoothes over the noise in the density field due to the discreteness
of particles.

\begin{figure}[t!]
\centering
\includegraphics[width=0.48\textwidth]{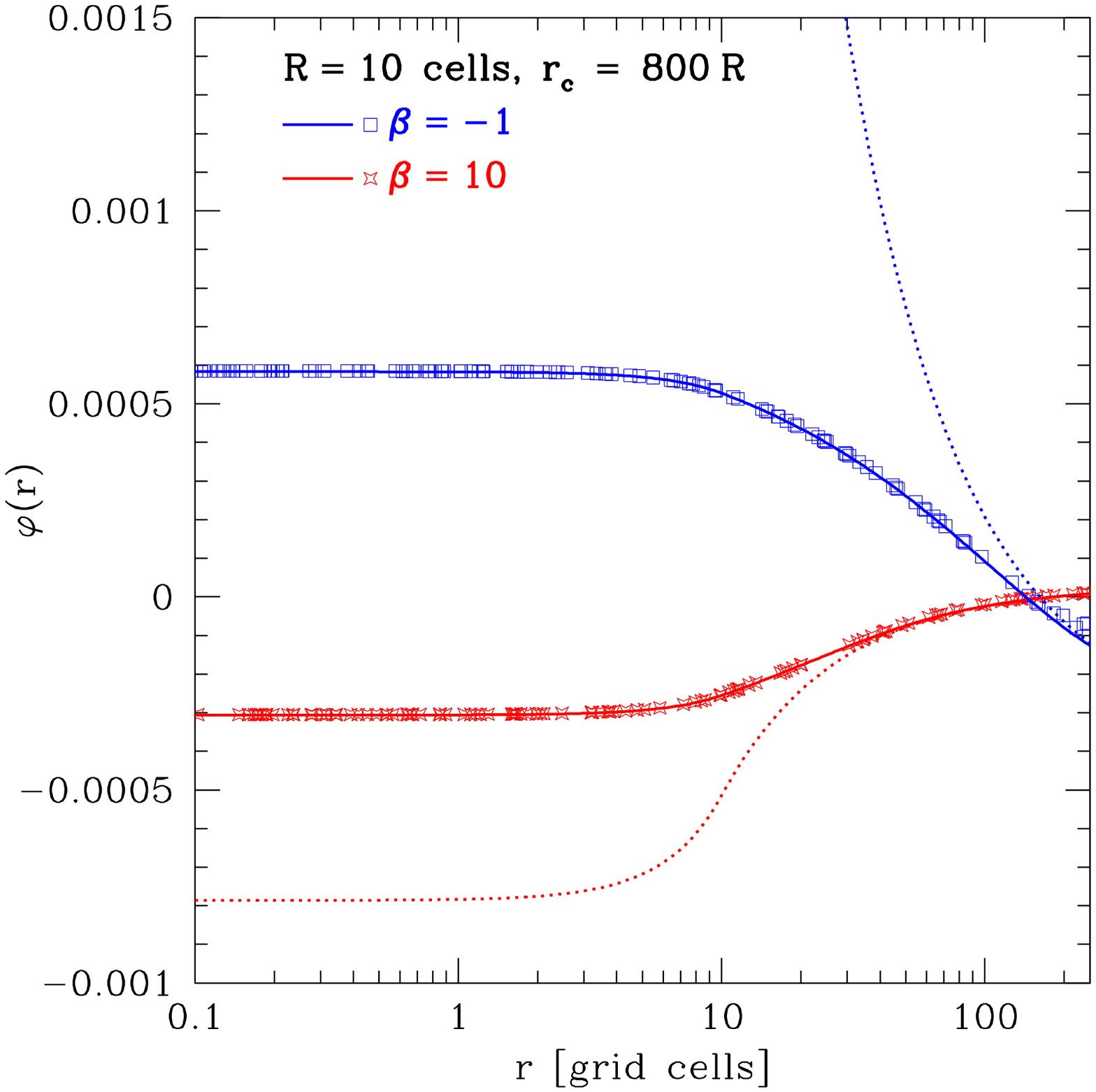}
\includegraphics[width=0.48\textwidth]{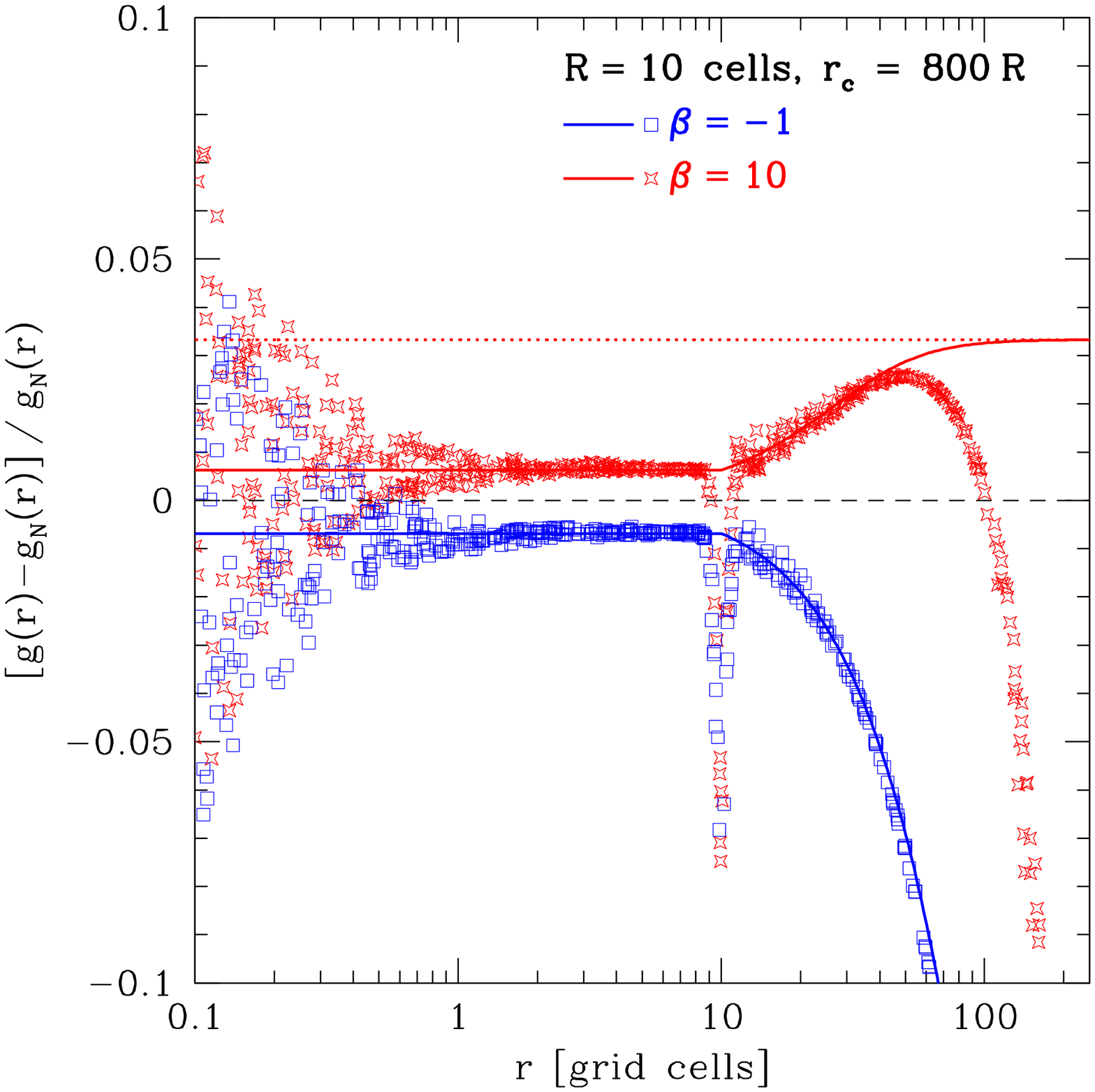}
\caption{{\small \textit{Left panel:} $\ph(r)$ measured in simulations
with a $512^3$ grid for a spherical mass for different values of $\beta$, 
where $r$ is the distance from
the center of the mass. $R=10$~grid cells, and $r_c = 800\,R$ is held fixed
(see text).
The solid lines show the full analytical solution for the spherical mass, while
dotted lines show the linearized solution (\refsec{philin}).
\textit{Right panel:} Relative deviation of the acceleration measured
in the spherical mass solution from the analytical Newtonian value vs. $r$, for the same
parameters as in the left panel. Again, solid lines show the full
exact solution, while dotted lines show the linearized solution.}
\label{fig:sphericalmass}}
\end{figure}

\section{Code tests}
\label{sec:tests}

In this section, we present different tests our code was put through
to benchmark its performance and limitations. We focus on tests applying
to the modified gravity sector. For the results of standard N-body code
tests for this code, see \cite{HPMpaper}.

\subsection{Spherical mass test}
\label{sec:sphericalmass}

In order to study how well the code reproduces the exact result
for a spherical mass in DGP, we start out with a grid with
$N_g^3$ grid cells and $N_P$ particles. We arbitrarily assume a box
size of $\Lbox = 200\Mpch$ and set $a=1$. All particles are moved
into a spherical mass of radius $R$ and uniform density, 
which then corresponds to
a mass of $M=\rho_{m0}\Lbox^3\approx 7\cdot 10^{17}\Msunh$
(assuming $\Om=0.3$).
Given the density field assigned from the particle positions, we then 
use the relaxation solver to solve for $\ph$, and measure the field
values and radial acceleration throughout the box.

For the purposes of this test, we can vary the two parameters
$\beta$ and $r_c$ in \refeq{phiQS} independently. $1/\beta$ determines
the strength of the additional force mediated by $\ph$ [e.g., \refeq{philin}], 
which is attractive for positive $\beta$ and repulsive for $\beta < 0$.
For given $\beta$, $r_c$ controls the Vainshtein radius, i.e. the scale
where the self-interactions of the $\ph$ field become important.

\reffig{sphericalmass} shows the field solution (left panel) and
relative deviation of the acceleration from the Newtonian value (right panel)
as a function of distance from the center, $r$,
for $N_g=512$, $R=10$~grid cells, corresponding to $3.9\Mpch$, and 
$r_c/R = 800 \approx 3100\Mpch$. The agreement with the analytical solution
(thick lines)
is generally very good, except at large radii, where the periodic
boundary conditions become important, and around $r = R$, where artifacts
of interpolating a sphere onto a cubic grid become visible. 
At distances of order the size of the grid cells, the force resolution
becomes worse leading to a growing scatter in the measured acceleration. 
Note that the analytical solution
necessarily assumes an isolated mass, since the superposition principle
cannot be used due to the non-linear $\ph$ field equation. For this reason,
we added an arbitrary zero-point to the $\ph$ field in order to match
the analytical solution in the left panel of \reffig{sphericalmass}.
Of course, such a zero-point does not affect any observable quantity,
such as the acceleration shown in the right panel.

We tested the field solution for several different values of $R$, $N_g$,
$\beta$, and $r_c$. Given the caveats pointed out, we found very good 
convergence to the analytical solution.
In particular, increasing $R$ reduces the interpolation artifacts
at $r=R$, while decreasing $R$ reduces the impact of the periodic
boundary conditions. Solutions for different values of $\beta$ 
are shown in both panels of \reffig{sphericalmass}. While the solid lines show
the exact solution to the full equation, \refeq{phiSph}, the dotted lines
show the linearized result, \refeq{philin}. Clearly, for the parameters chosen,
the effect of the non-linearity is significant for $r$ of a few $R$ or less.

\begin{figure}[t!]
\centering
\includegraphics[width=0.48\textwidth]{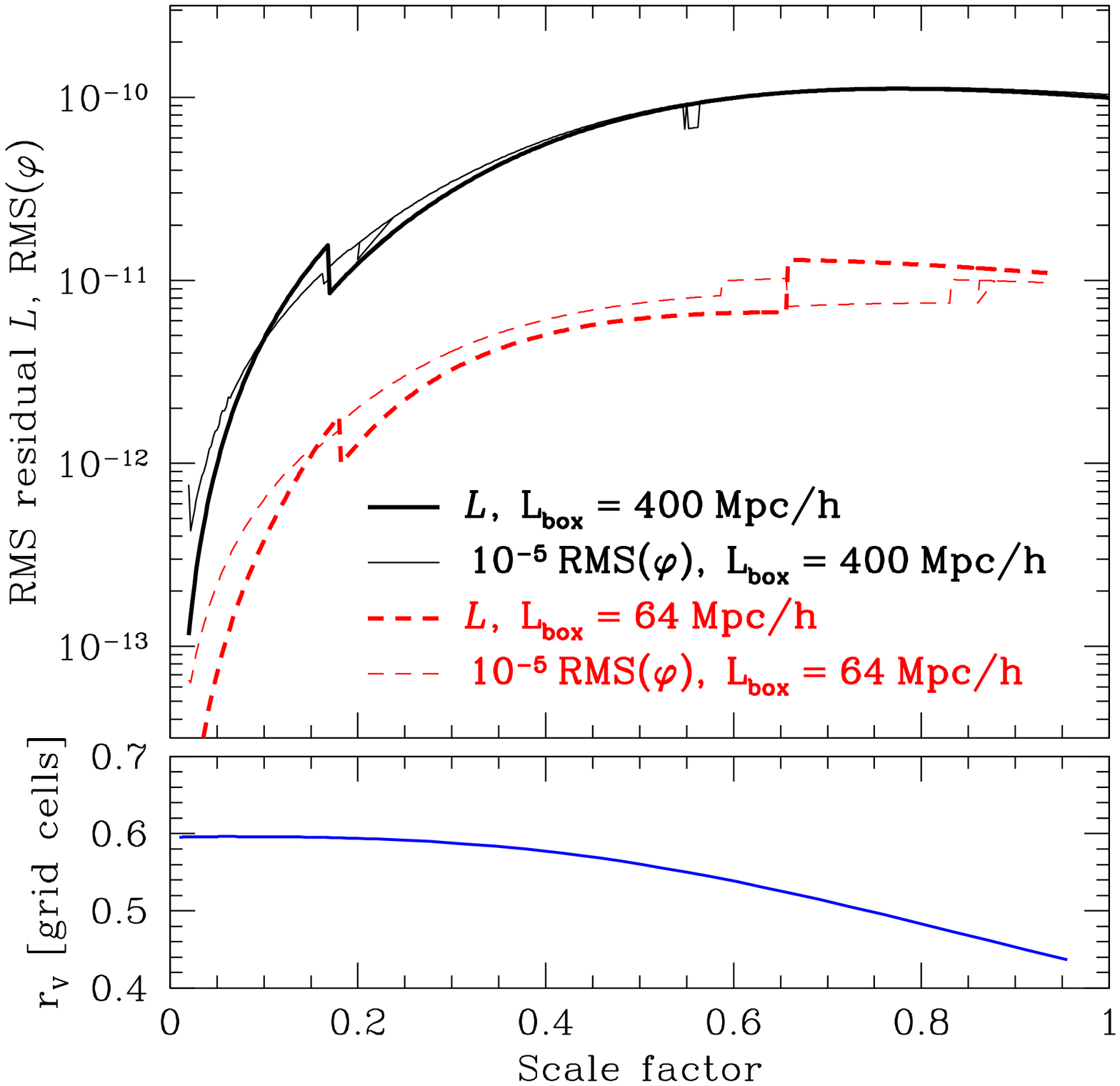}
\includegraphics[width=0.48\textwidth]{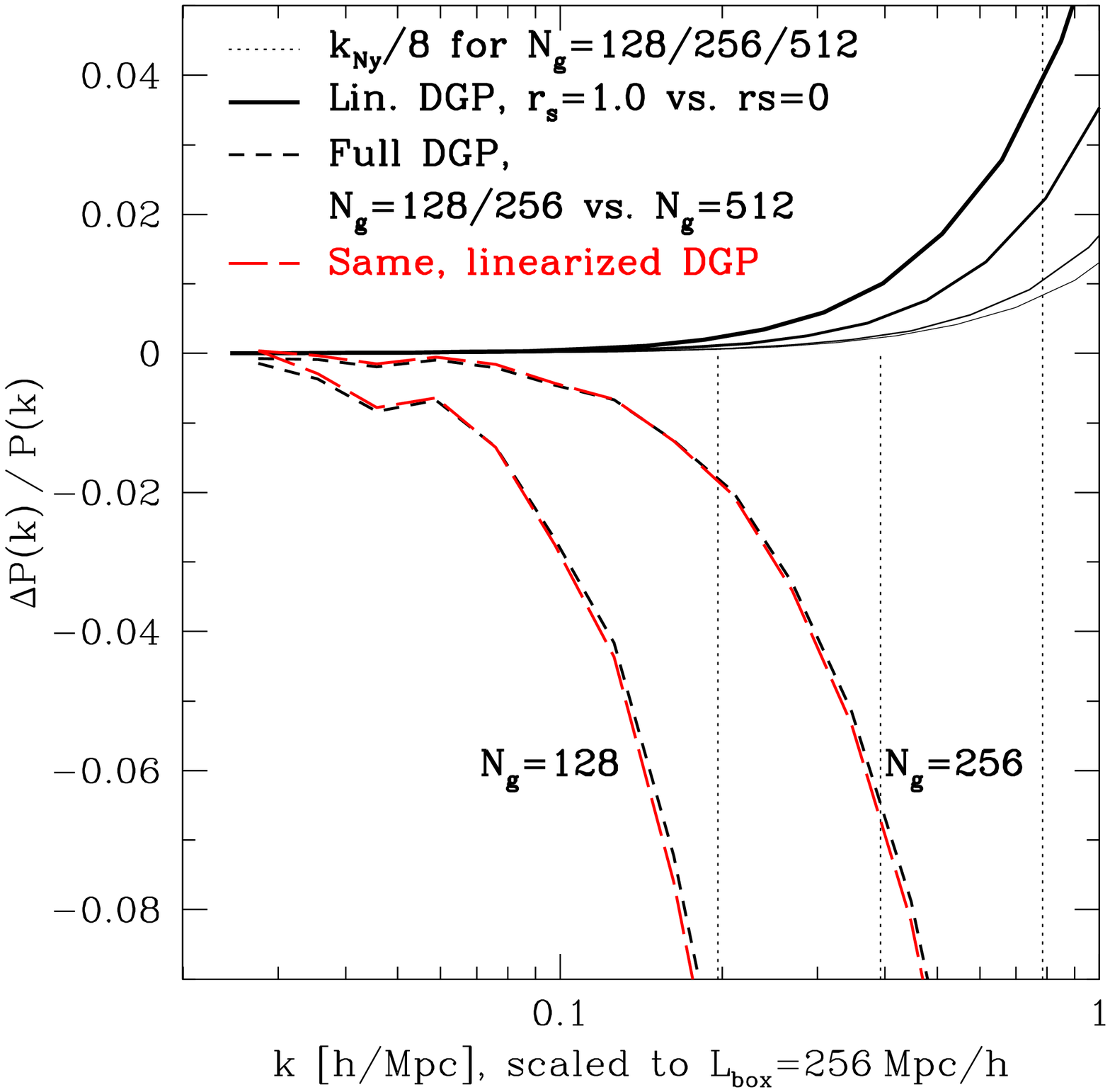}
\caption{{\small \textit{Left panel:} Dimensionless RMS residual 
$\L=\sqrt{\langle r^2\rangle}$ and RMS of the
$\ph$ field vs. $a$ in cosmological simulations (top), and Vainshtein
radius for a single particle in the simulations, in units of $r_{\rm cell}$
(bottom).
\textit{Right panel:} Relative deviation of linearized DGP power spectra
smoothed with $r_s/r_{\rm cell}=1.0$ from the unsmoothed simulations (solid lines,
for box sizes from $64\Mpch$ [thick] to 400$\Mpch$ [thin]), and
relative deviation of power spectra for lower resolution simulations
($N_g=128$, left dashed curves, $N_g=256$, right dashed curves).
Also shown as vertical dotted lines are the maximum wavenumbers considered
for each simulation, $k_{\rm max}=k_{\rm Ny}/8$.}
\label{fig:conv}}
\end{figure}

\subsection{Convergence and resolution tests}
\label{sec:convtest}

In most test cases we studied, the density field is sufficiently
smooth that the relaxation converges to the desired low tolerance.
However, for density
fields with considerable small-scale inhomogeneities, such as the
cosmological density field at late times, we find that this level
is not achievable. This is because the non-linearity of the $\ph$ equation
is important as soon as the overdensity $\delta$ becomes of order unity,
which in the N-body simulation corresponds to $\pm 1$ particle per grid
cell. Equivalently, the Vainshtein radius for a single particle
in our N-body simulation is of order the grid scale (see \reffig{conv},
bottom left). Thus, the $\ph$ field reacts much more strongly to noise due 
to the discreteness of particles than the Newtonian
potential, leading to increased residual errors of the approximate solution.

If we write \refeq{phiQS} in code units (see \refapp{code}) 
as $L(\ph)=f$, then for
an approximate solution $\ph_i$, $r \equiv r_{\rm cell}^2( L(\ph_i)-f)$ is 
the dimensionless {\it residual}, where $r_{\rm cell}=\Lbox/N_g$ is the size
of a grid cell. In the 
following, we use the RMS of the dimensionless residual, 
$\L \equiv \sqrt{\langle r^2 \rangle}$ as a benchmark for the convergence of 
the field solution.
We performed tests with sine wave density fields of various wavelengths
in order to determine what residual $\L$ is acceptable in our simulations
(see \refapp{sinetest}).
In case of a pure sine wave, the non-linearity in \refeq{phiQS} vanishes,
and the exact non-linear solution is identical to the linearized solution.
We found that residuals of $\L \lesssim 10^{-10}$ are safe, as for
residuals at that level, the errors in the
solution are negligible compared to the unavoidable truncation errors from taking
numerical derivatives on the grid.

In order to reduce the noise in the density field in the cosmological 
simulations and improve the solution to the acceptable level of $\L$, 
we increase the number of particles from $(N_g/2)^3$ 
to $N_g^3$, and smooth the density field entering the
r.h.s. of \refeq{phiQS} as described in \refsec{code}.
With these steps, the solution converges with a residual 
$\L \lesssim 10^{-10}$ for all box sizes (left panel of \reffig{conv}).
Note that the dimensionless RMS residuals never exceed a fraction of 
$\sim 10^{-5}$ of the RMS of the $\ph$ field solution.

Since the force resolution in our simulations is in any case limited to scales 
above $r_{\rm cell}$, a smoothing on the grid scale is not expected to degrade
the resolution of the simulations significantly. We checked this by running
linearized DGP simulations given by \refeq{philin} (\refsec{philin}) 
using the same smoothing of the r.h.s. of
\refeq{phiQS}, and comparing the resulting power spectrum to that of
linearized DGP simulations without smoothing, for the same initial conditions. 
The result is shown in the 
right panel of \reffig{conv}. As expected, the effect of smoothing
increases towards larger $k$. The smoothing effect on the matter power 
spectrum is positive, since the smoothing removes power in the $\ph$ field on the
smallest scales, and $\ph$ mediates a repulsive force. This leads
to an enhancement in the matter power spectrum.

The smoothing effect remains below 4\% for $k < k_{\rm max}=\kNy/8$,
which we adopt as the maximum wave number considered for each box,
where $\kNy = \pi\,N_g/\Lbox$ is the Nyquist frequency of the grid.
From our studies with different smoothing radii $r_s=0.8\dots 3.0$,
we found that the smoothing effects on full DGP simulations show a similar
$k$-dependence as those of the linearized DGP simulations, but are smaller 
by a factor of $\sim 0.6$. This is understandable since the $\ph$ field 
is suppressed in 
dense regions in the full simulations, reducing the effect of the smoothing.
Taking into account this factor, we correct the power spectra measured
in the full DGP simulations for the smoothing effects using the curves
shown in \reffig{conv} (right panel). Note that in any case these 
effects are at the level of few percent or less.

In order to assess the effect of the finite grid resolution, we also
performed simulations with $N_g=256$ and $N_g=128$, and $N_p=N_g^3$.
\reffig{conv} (right panel) also shows the deviation of the power spectrum
in these low-resolution simulations from the $N_g=512$ simulation with
the same initial conditions. The vertical lines indicate $k_{\rm max}=\kNy/8$ for
each case, i.e. the maximum $k$ we consider for each
simulation box. Below $\kNy/8$, the deviations are less than 10\% for either
$N_g=256$ or $N_g=128$.
Note also that the resolution effects are independent
of the type of simulation (DGP, linearized DGP, or GR). Hence, they
cancel when measuring the deviation of $P(k)$ between different simulation
types, and we expect this deviation to be accurate to the percent level for
the range in $k$ we consider.

\begin{figure}[t!]
\centering
\includegraphics[width=0.48\textwidth]{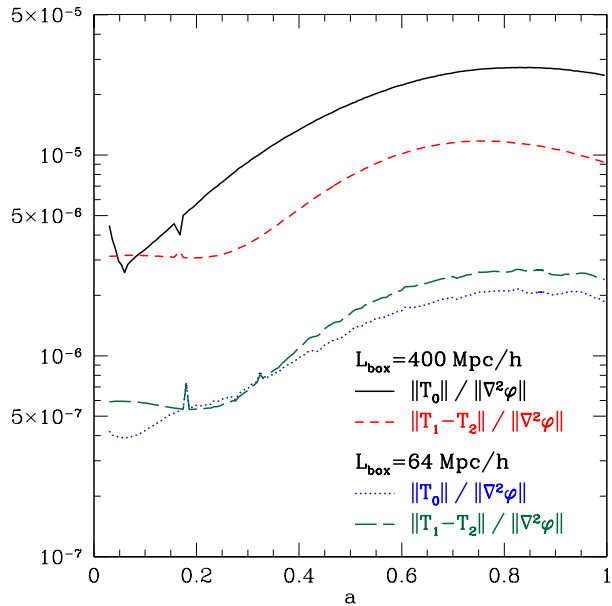}
\caption{{\small The time derivative terms \refeqs{T0}{T2} of $\ph$ measured in the 
simulations relative to the spatial Laplacian, as a function of scale factor
$a$ for our largest and smallest simulation boxes. The time derivatives always
remain $5-6$ orders of magnitude below the spatial derivatives.}
\label{fig:qstest}}
\end{figure}
\subsection{Quasi-static approximation}
\label{sec:qstest}

In solving for the brane bending mode $\ph$, we have assumed that
all time-derivative terms in the full equation of motion of $\ph$
(in terms of physical coordinates):
\be
\Box \varphi + \frac{r_c^2}{3\beta} \left [ (\Box\varphi)^2 - (\nabla_\mu\nabla_\nu\varphi)^2 \right ] = \frac{8\pi\,G}{3\beta} \delta\rho,
\ee
can be neglected with respect to the spatial derivatives in \refeq{phiQS}. 
Up to second order, there are three
time derivative terms which have been neglected:
\begin{eqnarray}
T_0 &=& a^2 \partial_{tt}\varphi \label{eq:T0}\\
T_1 &=& \frac{2 r_c^2}{3\beta} \partial_{tt}\varphi \cdot \nabla^2\varphi \label{eq:T1} \\
T_2 &=& \frac{2 r_c^2}{3\beta} \left ( \partial_t\nabla_i\, \varphi \right )^2.\label{eq:T2}
\end{eqnarray}
$T_1$ and $T_2$ appear as $T_1-T_2$ in the equations of motion.
While we cannot rigorously prove that the quasi-static approximation holds,
we can perform a consistency test by measuring the terms \refeqs{T0}{T2}
in the simulations, and checking whether they are indeed small.
The time derivatives of $\ph$ are calculated in each grid cell using
the two neighboring time steps. We then take the RMS of each term over the grid, 
$\parallel T_i\parallel = \sqrt{\langle T_i^2\rangle}$, and compare
with the RMS of the spatial Laplacian, $\parallel \nabla^2\ph\parallel$.
\reffig{qstest} presents $T_0$ and $T_1-T_2$ relative to the spatial derivatives
for our largest and smallest box sizes, and shows that they are well under
control. The relative magnitude of
all time derivatives are of order a few $10^{-5}$ or less. This is in keeping
with the expectation that $\partial\ph/\partial t \sim H\,\ph$,
and $k/a \gg H$ in our simulations.
We conclude
that the quasi-static approach is a self-consistent approximation.

\section{Cosmological simulations}
\label{sec:sims}

\begin{table}[b!]
\begin{minipage}[t]{0.48\textwidth}
\caption{Cosmological parameters. \label{tab:params}} 
\begin{center}
  \leavevmode
  \begin{tabular}{l|l}
\hline
$\Om$ & \  0.258 \\
$\Orc$ & \  0.138 \\
$H_0$ & \  66.0 $\rm km/s/Mpc$\\
\hline
$100\,\Omega_b\,h^2$ & \  2.37 \\
$\Omega_c\, h^2$ & \  0.0888 \\
$\tau$ & \  0.0954 \\
$n_s$ & \  0.998 \\
$A_s$ ($k=0.05\,{\rm Mpc}^{-1}$) & \  $2.016\: 10^{-9}$ \\
\hline\hline
$\sigma_8(\Lambda\rm CDM)$ & 0.6566\footnote{Linear power spectrum normalization today
of a $\Lambda$CDM model with the same primordial normalization.} \\
\hline
\end{tabular}
\end{center}
\end{minipage}
\hfill
\begin{minipage}[t]{0.48\textwidth}
\caption{Simulation type and number of runs per box size. \label{tab:runs}} 
\begin{center}
  \leavevmode
  \begin{tabular}{c|c|c c c c}
\multicolumn{2}{c|}{ } & \multicolumn{3}{|c}{$L_{\rm box}$ [$\Mpch$]} \\ 
  \cline{3-6} 
\multicolumn{2}{c|}{ } &\ \ $400$\ \ & $256 $\ \ \  & $128$\ \ \  & $64 $\ \ \  \\
\hline
\# of\ \ & QCDM & 6 & 6 & 6 & 6\\
boxes\ \ & Lin. DGP & 6 & 6 & 6 & 6\\
     & Full DGP & 6 & 6 & 6 & 6\\
\hline
\multicolumn{2}{c|}{$\kmax=\kNy/8$ [$h/\rm Mpc$]} & 0.50 & 0.79 & 1.57 & 3.14 \\
\hline
\multicolumn{2}{c|}{$r_{\rm cell}$ [${\rm Mpc}/h$]} & 0.78 & 0.50 & 0.25 & 0.13 \\
\hline
\multicolumn{2}{c|}{$M_{\rm min}$ [$10^{12}\Msunh$]} & 219 & 57.3 & 7.17 & 0.90 \\
\hline
\end{tabular}
\end{center}
\end{minipage}
\end{table}
We performed a suite of cosmological simulations for three different
types of gravity: unmodified GR, corresponding to a smooth dark energy
model with the same expansion history as DGP, referred to as {\it QCDM};
\textit{linearized DGP} gravity, using the linearized $\ph$ equation
of motion [\refeq{philin}], corresponding to a time-dependent
rescaling of Newton's constant; and \textit{full DGP}, solving for the
full non-linear $\ph$ solution [\refeq{phiQS}]. Comparing linearized
DGP with full DGP allows us to study the Vainshtein effect in a cosmological
setting.

The cosmological parameters are those of the best-fit flat
self-accelerating DGP model to WMAP 5yr data \cite{FangEtal08} and
are summarized in \reftab{params}. We generated
the initial conditions at $a=0.02$ ($z=49$) from a modified transfer function 
output of CAMB \cite{CAMB} for a flat $\Lambda$CDM model with 
$\Omega_\Lambda=1-\Omega_m$. The $\Lambda$CDM transfer function was corrected
for small early-time modified gravity effects in the DGP model using the PPF 
approach, as detailed in \refapp{ICcorr}. 

The simulations were run with $N_g=512$ grid cells on a side, and $N_p=512^3$
particles, i.e. one particle per grid cell, to reduce the shot noise in 
the density field (see \refsec{convtest}). We performed six simulations
each for four different box sizes, from $\Lbox=64\Mpch$ up to 
$\Lbox=400\Mpch$ (\reftab{runs}). On an 8-core machine, the QCDM and
linearized DGP runs require $\sim$10h of computing time, while the
full DGP simulations require about 300h.

\begin{figure}[t!]
\centering
\includegraphics[width=0.48\textwidth]{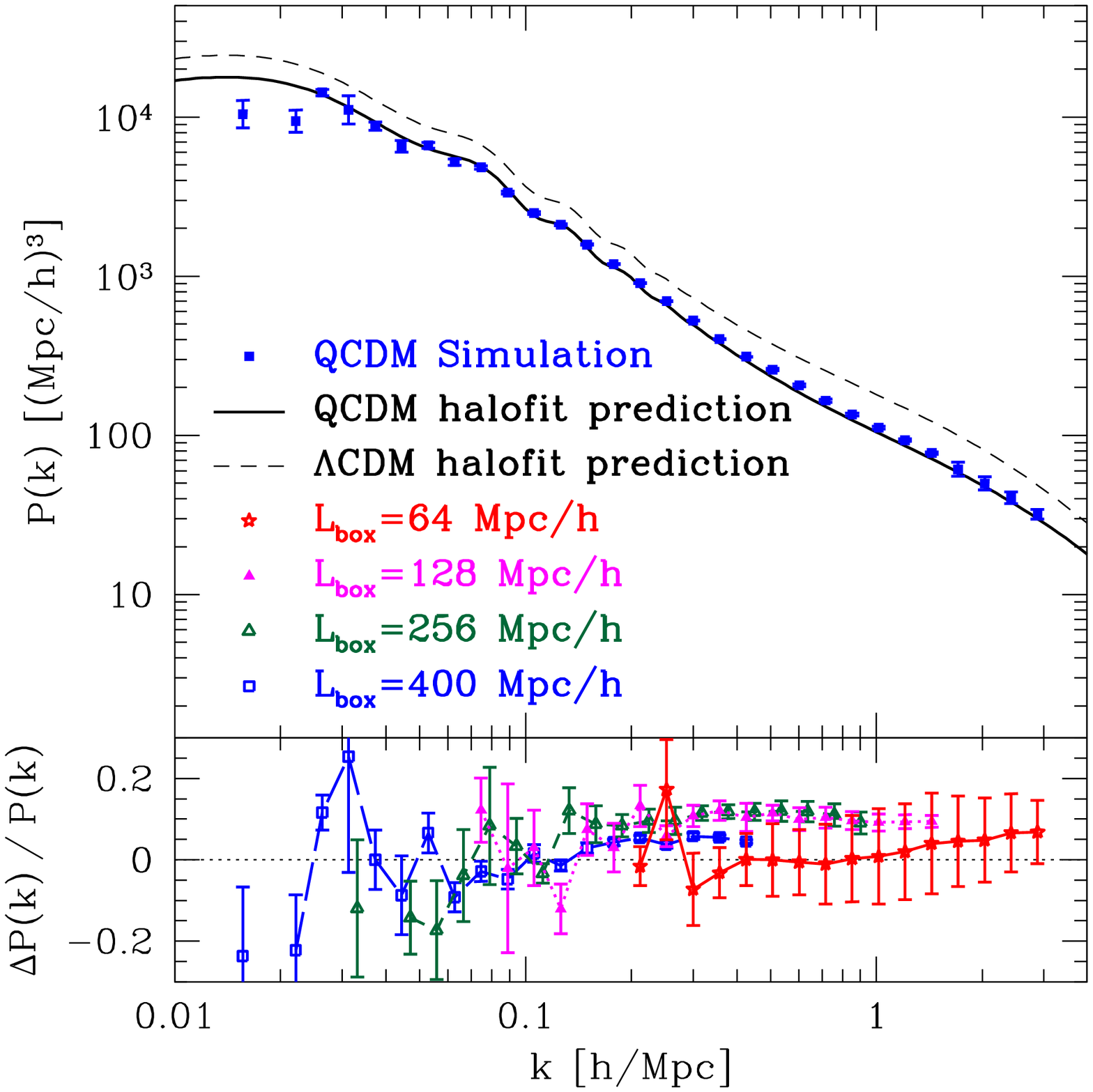}
\includegraphics[width=0.48\textwidth]{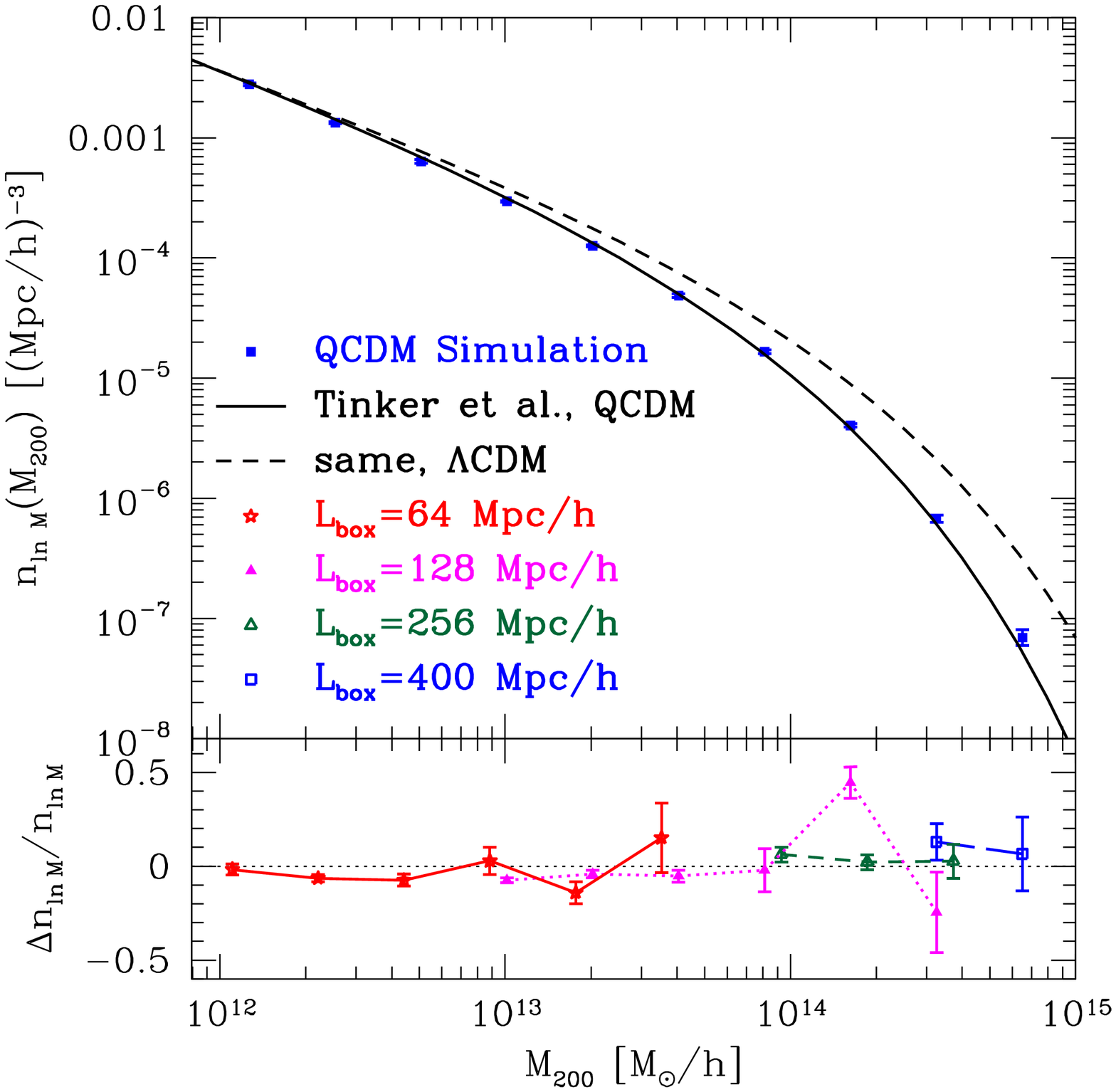}
\caption{{\small \textit{Left panel:} Power spectrum measured in the QCDM effective dark
energy cosmology simulations (top), and \texttt{halofit} predictions.
The lower panel shows the deviations from the \texttt{halofit} prediction
separately for each simulation box size. 
\textit{Right panel:} Halo mass function $dn/d\ln M_{200}$ measured in
the QCDM simulations (points) and the mass function prediction for QCDM after
\cite{TinkerEtal}. Also shown is the mass function for a $\Lambda$CDM
model with the same primordial power spectrum.
The lower panel shows the deviation from the prediction separately for
each box size.}
\label{fig:PkQCDM}}
\end{figure}

\reffig{PkQCDM} (left panel) shows the combined power spectrum from all box sizes
for the QCDM simulations, including bootstrap errors. Also shown
is the non-linear power spectrum for QCDM, calculated from the
linear power spectrum using the \texttt{halofit} procedure \cite{halofit}.
The power spectrum
measured in each box is used up to $\kmax=\kNy/8$ (see \reftab{runs}),
and different boxes are combined weighting by volume. The lower panel
of \reffig{PkQCDM} (left) shows the power spectrum relative to the \texttt{halofit} 
prediction, measured separately for each box size. The power
spectra measured in different boxes clearly agree within the errors, and the deviations
from the \texttt{halofit} prediction are within the accuracy ($\sim20$\%) expected
from this fitting procedure (especially given the significant departures 
from $\Lambda$CDM of the simulated expansion history).

\reffig{PkQCDM} also shows the non-linear power spectrum for a flat $\Lambda$CDM
cosmology fixed to the same $\Om$, $h$, and primordial normalization, 
corresponding to a linear normalization of $\sigma_8=0.66$ today. Note
that due to the different expansion history and the earlier onset of
acceleration in DGP, the linear growth factor is suppressed by $\sim15$\% 
in QCDM relative
to $\Lambda$CDM. The repulsive brane-bending mode in DGP suppresses growth
further, leading to a suppression \textit{in the linear regime} of $\sim 21$\%
relative to $\Lambda$CDM.
In the following, we will always compare the DGP results
with QCDM, so that the expansion history effects are taken out, and
all deviations are strictly due to modifications of gravity.

\begin{figure}[t!]
\centering
\includegraphics[width=0.48\textwidth]{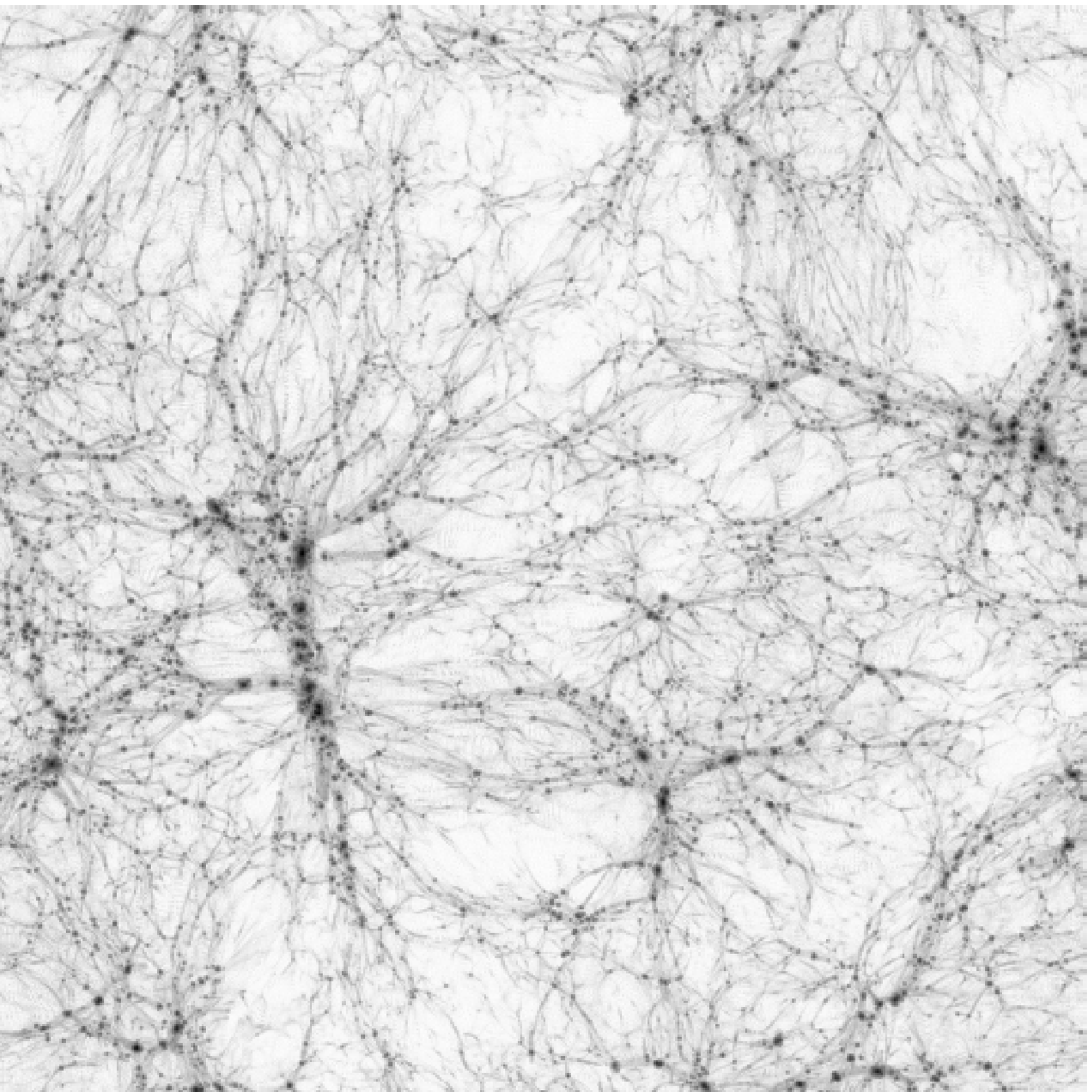}
\includegraphics[width=0.48\textwidth]{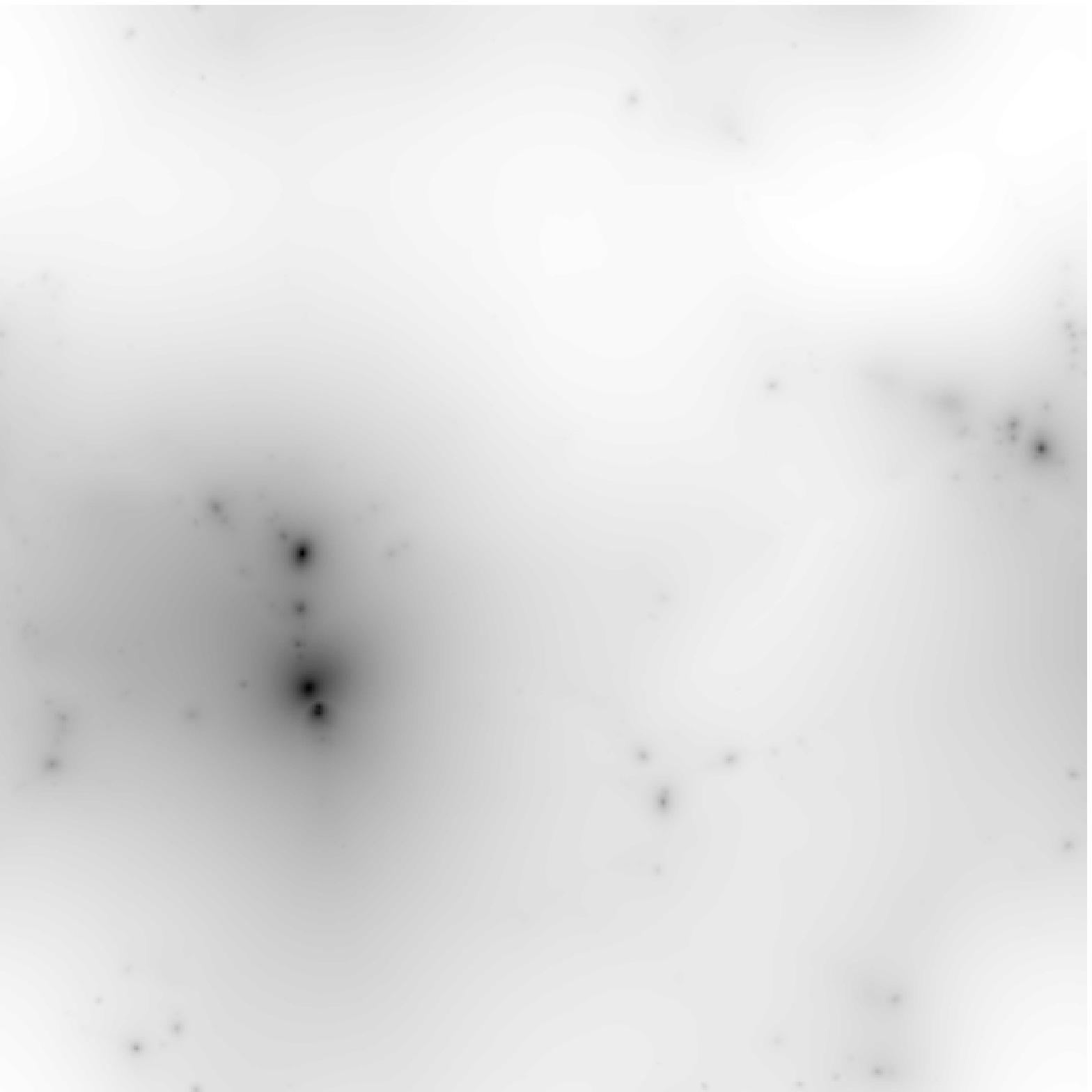}
\includegraphics[width=0.48\textwidth]{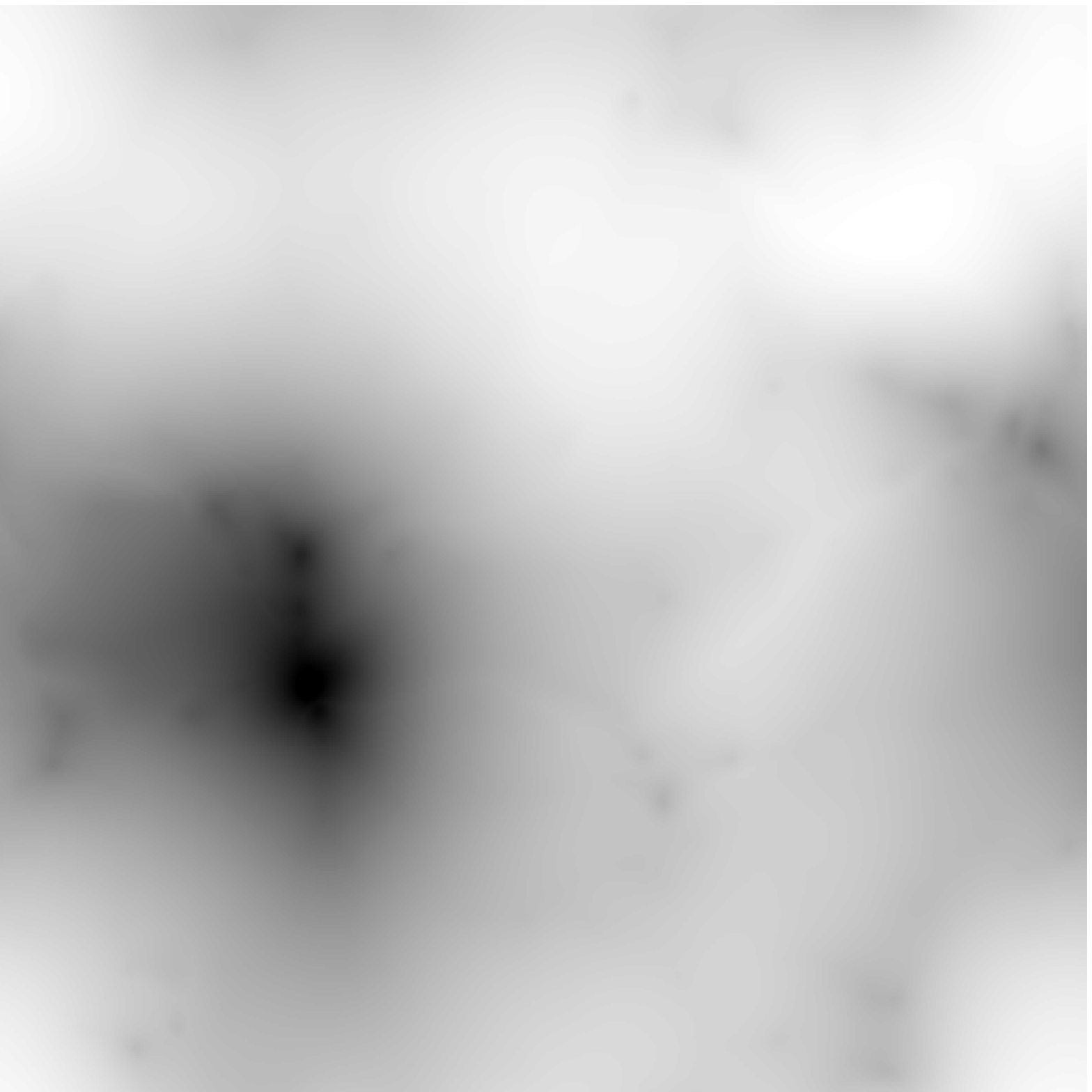}
\includegraphics[width=0.48\textwidth]{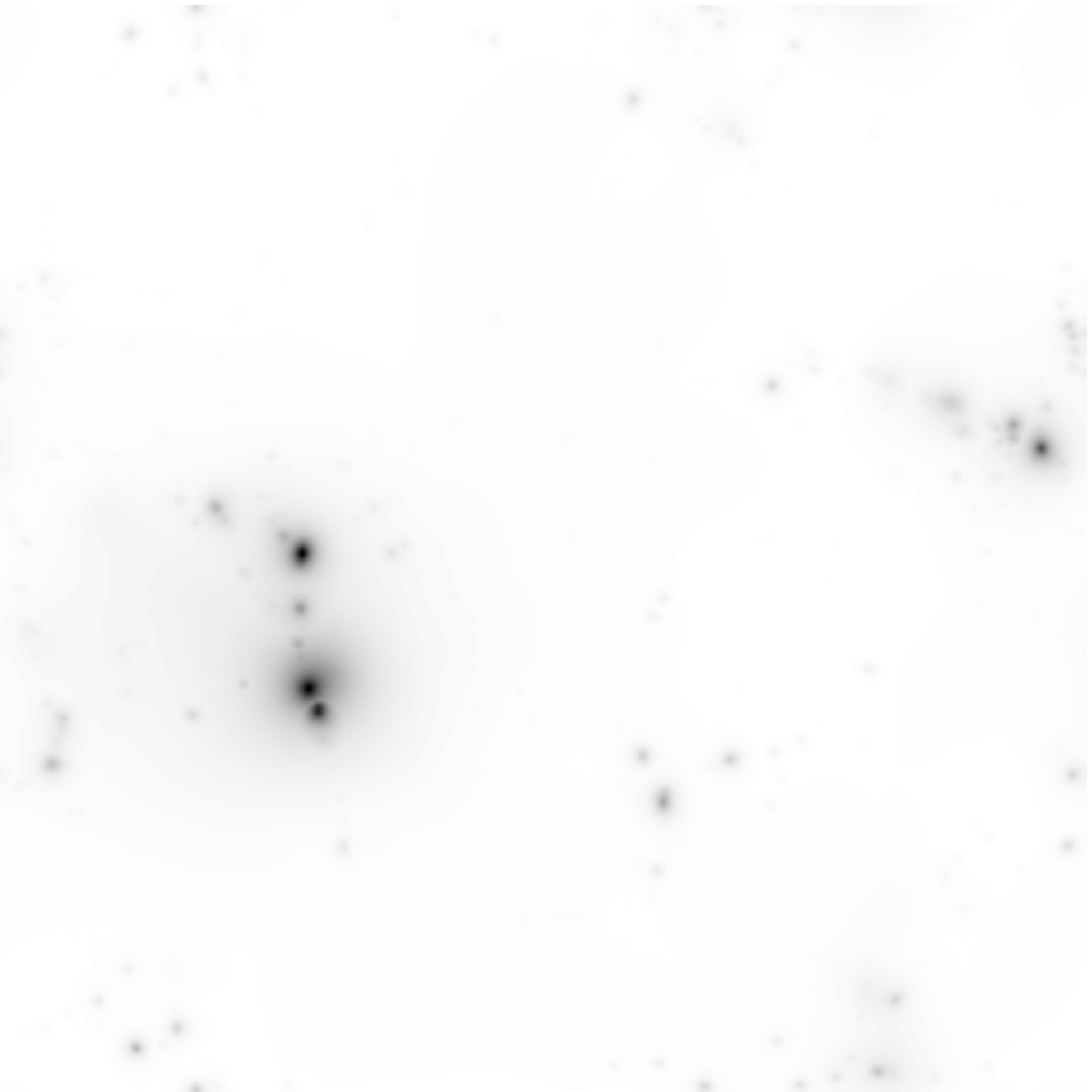}
\rput(-10,-15){\Large $\Psi$}
\caption{{\small Slices through a full DGP simulation with $\Lbox=64\Mpch$
at $z=0$. 
\textit{Top left}: density in logarithmic scale;
\textit{top right}: dynamical potential $\Psi$;
\textit{bottom left}: brane-bending mode $\ph$;
\textit{bottom right}: non-linear suppression of $\ph$ field: 
$\ph_{\rm NL} \equiv \ph - \ph_{L}$ (see text).}
\label{fig:slice}}
\end{figure}
We have also measured the mass function of dark matter halos in our
simulations. The spherical overdensity halo-finding and 
combination of different simulation
boxes was done as described in \cite{HPMhalopaper}. As our grid resolution
is the same as in the simulations analyzed in \cite{HPMhalopaper}, while
we have a factor of 8 more particles, we conservatively increase the minimum 
required
number of particles from 800 to 6400. The corresponding mass thresholds
are given in \reftab{runs}. We define the halo mass $M_{200}$ as the
mass enclosed within a sphere of radius $R_{200}$, so that the
average density within the sphere is 200 times the average matter density in 
the universe, $\bar\rho_m=\Omega_m\rho_{\rm crit}$.

\reffig{PkQCDM} (right panel) shows the mass function,
$n_{\ln M}\equiv dn/d\ln M_{200}$,
measured in QCDM simulations, in comparison with the prediction calculated 
from the linear QCDM power spectrum using the fitting formula from
Tinker et al.\cite{TinkerEtal}. The lower panel shows the relative
deviations from the prediction, separately for each simulation box.
Within the mass range accessible with
our simulations, $M_{200} \sim 10^{12}-10^{15}\Msunh$, the agreement with the 
prediction and among different boxes is good. We also show the predicted
mass function
for a $\Lambda$CDM model with the same initial power spectrum in 
\reffig{PkQCDM}. Apparently, the number of halos above $\sim 10^{14}\Msunh$
is significantly reduced due to the suppressed growth in QCDM.

\section{Results}
\label{sec:res}

In order to get a visual impression of some of the physics in DGP N-body
simulations, we show slices through one simulation box at $z=0$ 
($\Lbox=64\Mpch$) in \reffig{slice}. The slices are 64 cells thick, and
for each pixel we take the maximum absolute values of each quantity 
over the thickness of 
the slice, for better visibility. The density field (top left; in 
logarithmic scale) is 
difficult to distinguish visually from the QCDM result for the same run,
as the effects on the power spectrum are at the $\sim15$\% level 
(\refsec{Pk}). The top right panel in \reffig{slice} shows the dynamical 
potential $\Psi$, exhibiting the potential wells of massive collapsed 
structures. The brane-bending mode $\ph$ is shown in the lower left panel. 
On linear scales, $\ph \approx \ph_L$ is proportional to the potential $\Psi$ 
(\refsec{philin}), but evidently does not follow the
potential within deeper potential wells, making it appear smoother.
To make this more clear, we show the difference of the $\ph$ solution
from the linear value [\refeq{philin}], $\ph_{\rm NL}\equiv\ph-\ph_{L}$,
in the lower right panel of \reffig{slice}: $\ph$ is suppressed ($\ph_{\rm NL} < 0$)
in overdense regions, showing the Vainshtein effect at work in a 
cosmological setting. Note that quite low-mass structures which are
not conspicuous in the potential $\Psi$ already lead to a suppression
of $\ph$. See \refsec{profiles} for a discussion of the Vainshtein effect
in dark matter halos.

\begin{figure}[t!]
\centering
\includegraphics[width=0.48\textwidth]{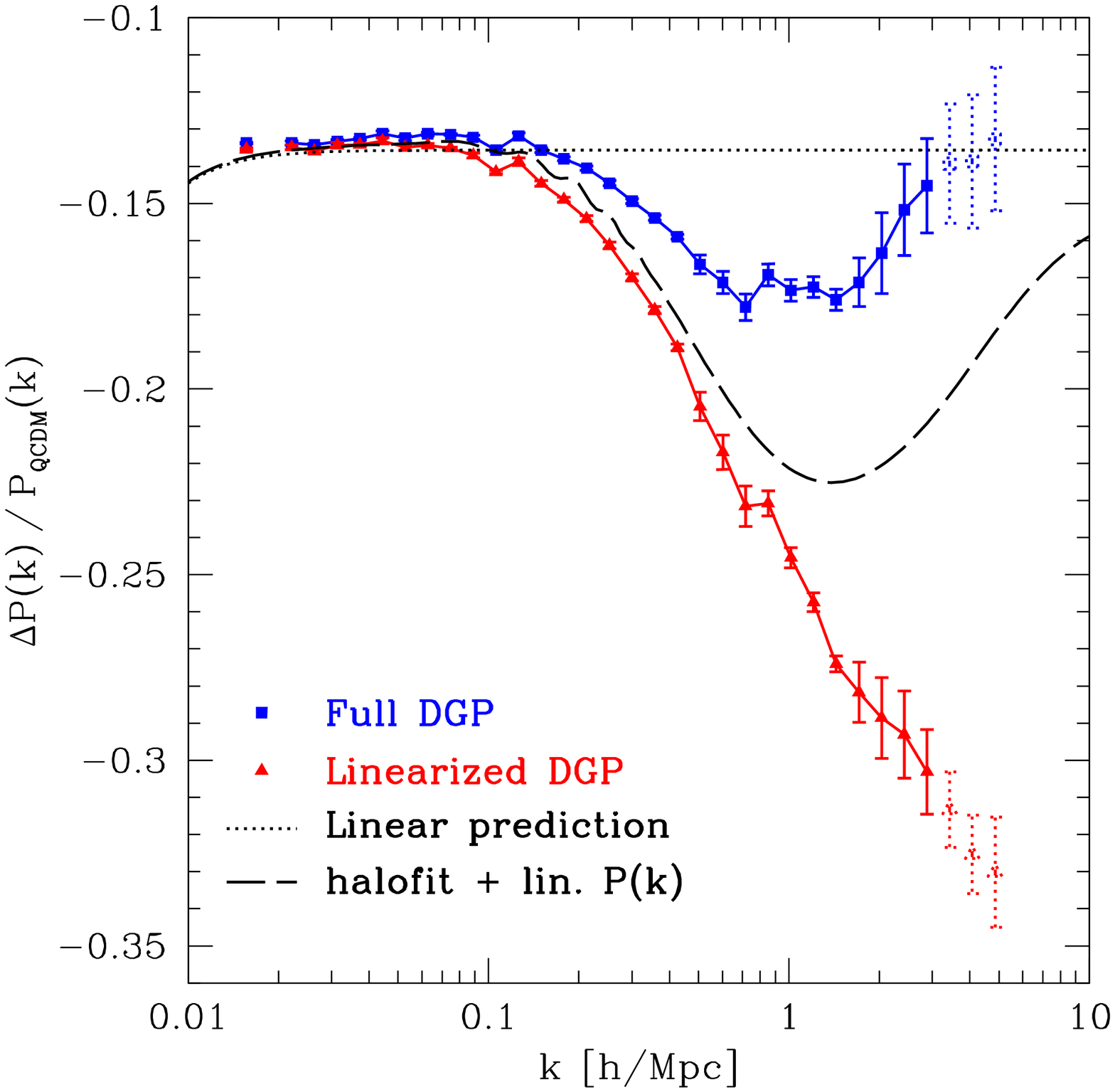}
\includegraphics[width=0.48\textwidth]{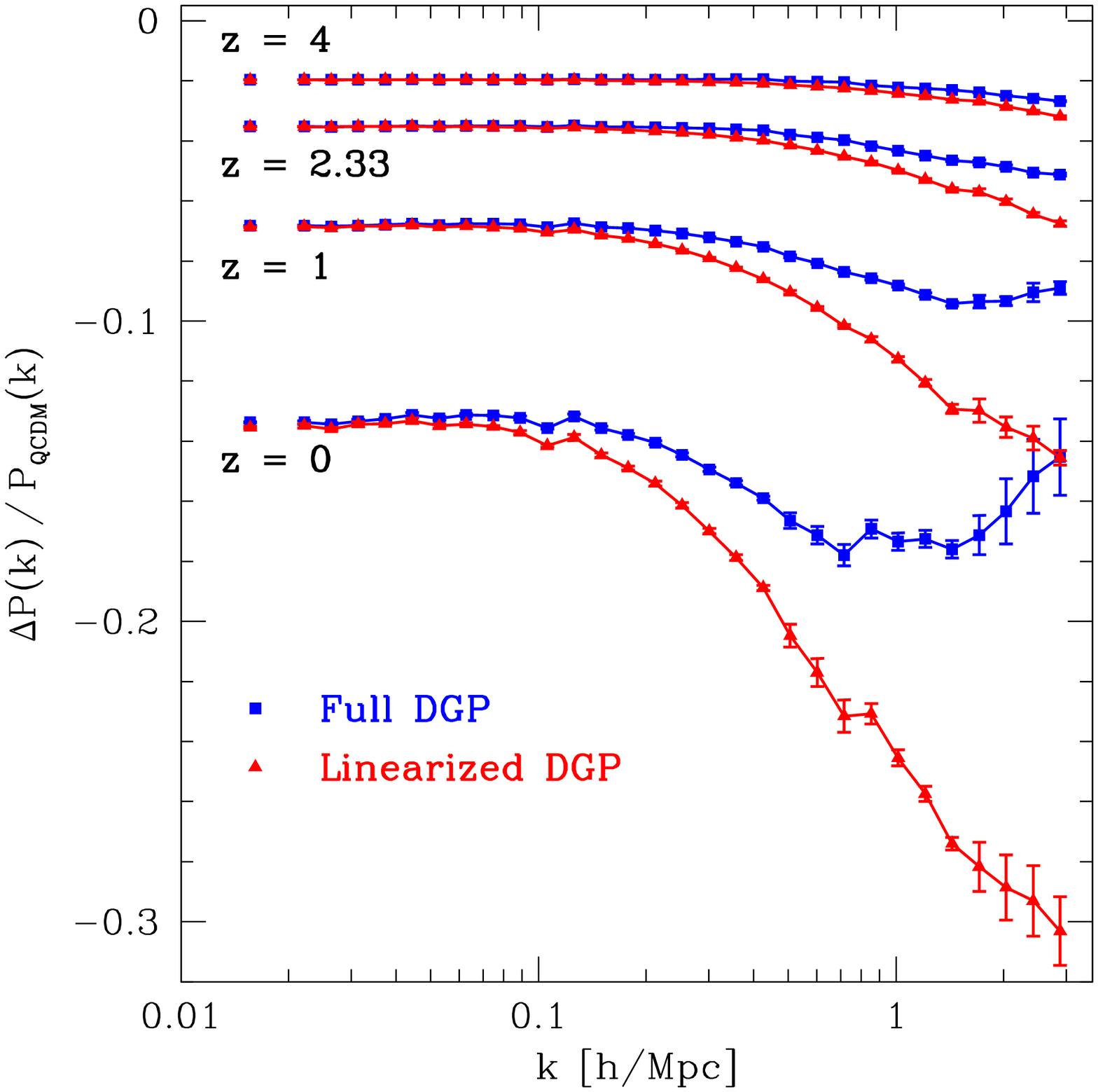}
\caption{{\small \textit{Left panel:} Relative power spectrum deviation of full and linearized DGP
simulations from GR (QCDM) at $z=0$. The dashed points at high $k$ extend the
range from $\kNy/8$ to $\kNy/4$ and are only intended to be indicative of 
the trend. The dotted line shows the deviation expected
in linear theory, while the dashed line indicates the deviation predicted
using linear theory as input for {\tt halofit}.
\textit{Right panel:} Same as the left panel, but measured in the
simulations at different redshifts.}
\label{fig:PkDGP}}
\end{figure}

\subsection{Power spectrum}
\label{sec:Pk}

\reffig{PkDGP} (left panel) shows the relative deviation of the linearized and full DGP
power spectra from the QCDM result at redshift 0. The error bars are
obtained from the 6 separate runs using a bootstrap procedure.
By comparing simulation runs with the same
initial conditions, and then averaging the deviations, most of the cosmic 
variance cancels out and we are able to obtain significantly less scatter.
Also shown is the predicted deviation in the {\it linear} power spectrum.
Note first that in the self-accelerated DGP branch, the scalar field
mediates a {\it repulsive} force, leading to a suppression of the growth
of structure.
Due to the scale-invariant modification of gravity in linearized
quasi-static DGP [\refeq{philin}], the predicted linear deviation is also 
scale-independent. On linear scales, $k\lesssim 0.1\iMpch$, both linearized 
and full DGP simulations
agree well with the linear prediction. 

Apparently, the full DGP result departs
significantly from linearized DGP on quasi-linear to non-linear
scales. The Vainshtein effect begins to operate wherever overdensities
become of order unity, and suppresses the deviation from GR. Note that
small effects of the non-linear $\ph$ equation can already be seen on
quite large scales, $k \sim 0.05\iMpch$ corresponding to $r\sim 60\Mpch$,
not far from the acoustic features in the matter power spectrum. While
we do not expect dramatic effects on cosmological parameter constraints
from BAO, a modeling of these non-linear effects will be necessary for 
precision constraints in the context of DGP and similar braneworld models.

We also
show the deviation of $P_{\rm NL,DGP}$ from $P_{\rm NL,QCDM}$, where
$P_{\rm NL}$ is calculated from the corresponding linear power spectrum
using the standard {\tt halofit} prescription. {\tt halofit} describes the 
linearized DGP power spectrum reasonably well up to $k \lesssim 0.4\iMpch$,
in agreement with the findings of \cite{LaszloBean},
while it follows neither the linearized nor full DGP at higher $k$.
We have not explored whether phenomenological modifications of {\tt halofit}
allow for an improvement of the fit.

The dashed points at high $k$ extend the range in frequency up to $\kNy/4$
of our smallest box, $64\Mpch$. While resolution effects are still expected
to cancel out to first order in this representation, these points only
serve to indicate that the trends seen for full and linear DGP at lower
$k$ continue towards smaller scales.

The right panel of \reffig{PkDGP} shows the evolution of the modified
gravity effects on the power spectrum as function of redshift. On large
scales $k\lesssim0.1\iMpch$, the deviation is almost scale-free and evolves as predicted by
linear theory. At earlier
times, the density field is non-linear only on smaller scales. Hence,
the Vainshtein effect becomes visible in the power spectrum only at 
higher $k$. However, it does affect the power spectrum deviation significantly
already at $z>1$.

\begin{figure}[t!]
\centering
\includegraphics[width=0.48\textwidth]{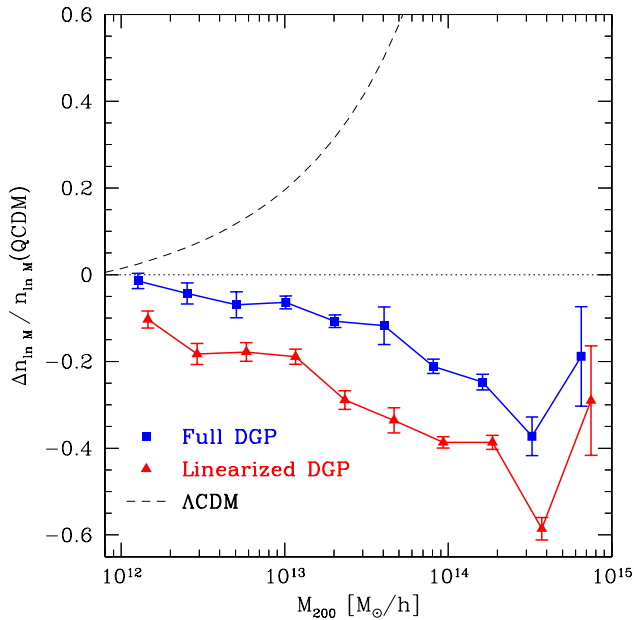}
\caption{{\small 
Relative deviation of the halo mass function $n_{\ln M}\equiv dn/d\ln M_{200}$ measured in
full and linearized DGP simulations from GR (QCDM) at $z=0$ (points). The 
dashed line shows the relative deviation of the predicted $\Lambda$CDM
mass function (for the same $\Omega_m$, $h$) from QCDM (see \reffig{PkQCDM}).
}
\label{fig:dndmDGP}}
\end{figure}

\subsection{Halo mass function}
\label{sec:massfct}

The abundance of massive dark matter halos is a sensitive probe
of the growth of structure in the Universe 
\cite{WhiteEtal93,EkeEtal98,BorganiEtal01,Vikhlinin}.
In particular, the number of the most massive halos which host galaxy
clusters depends exponentially on the amplitude of the matter power
spectrum. \reffig{dndmDGP} shows the effect of the modification of
gravity in DGP on the halo mass function, relative to the QCDM effective
dark energy cosmology. We investigated the effect of the smoothing
used in the full DGP simulations by comparing linearized DGP simulations
with and without smoothing, as done for the power spectrum (\refsec{convtest}).
Above our rather conservative mass threshold for each box, we only found a small 
effect of the smoothing on halo masses in the linearized DGP simulations.
Since the $\ph$ field is in any case suppressed within halos due to
the Vainshtein mechanism, we expect the smoothing effects to be even smaller
for the full DGP simulations.

As expected, the full DGP simulations are somewhat
closer to QCDM than the linearized DGP case in \reffig{dndmDGP}. 
The suppression in the mass
function is significant for $M_{200} \gtrsim 10^{13}\Msunh$, reaching
more than 30\% for massive cluster-size halos. The suppression is
smaller for lower-mass halos. This is presumably because these halos
formed earlier in cosmic history, when the modified gravity effects of 
DGP were significantly smaller.

Note that when compared to $\Lambda$CDM, this suppression comes in 
addition to the larger suppression of the mass
function due to the expansion history of DGP (see dashed line in
\reffig{dndmDGP}). However, we expect the effect of the DGP modification
of gravity to be fairly independent of the expansion history. In particular,
we expect a similar effect on the halo mass function in generalized
braneworld models which exhibit an expansion history close to $\Lambda$CDM
\cite{deRham,KW}. Hence, the mass function is expected to be a
sensitive probe for braneworld modified gravity models, as is the case
for $f(R)$ gravity \cite{HPMhalopaper}. Note that for the normal branch
of braneworld gravity, the sign of the effect in \reffig{dndmDGP}
will be reversed, leading to an enhancement of the number of massive halos.

\subsection{Brane bending mode and Vainshtein effect}
\label{sec:profiles}

\begin{figure}[t!]
\centering
\includegraphics[width=0.48\textwidth]{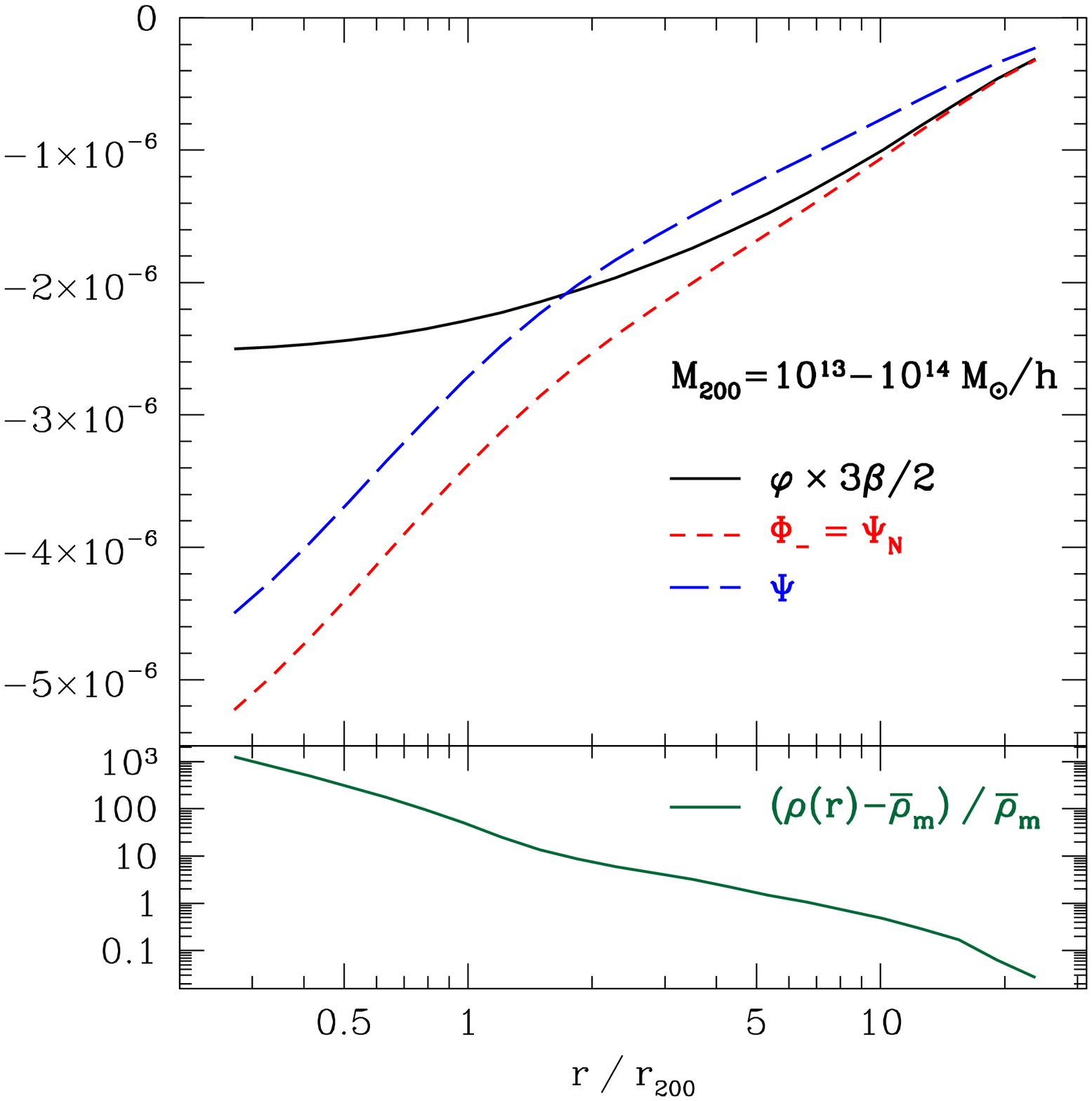}
\includegraphics[width=0.48\textwidth]{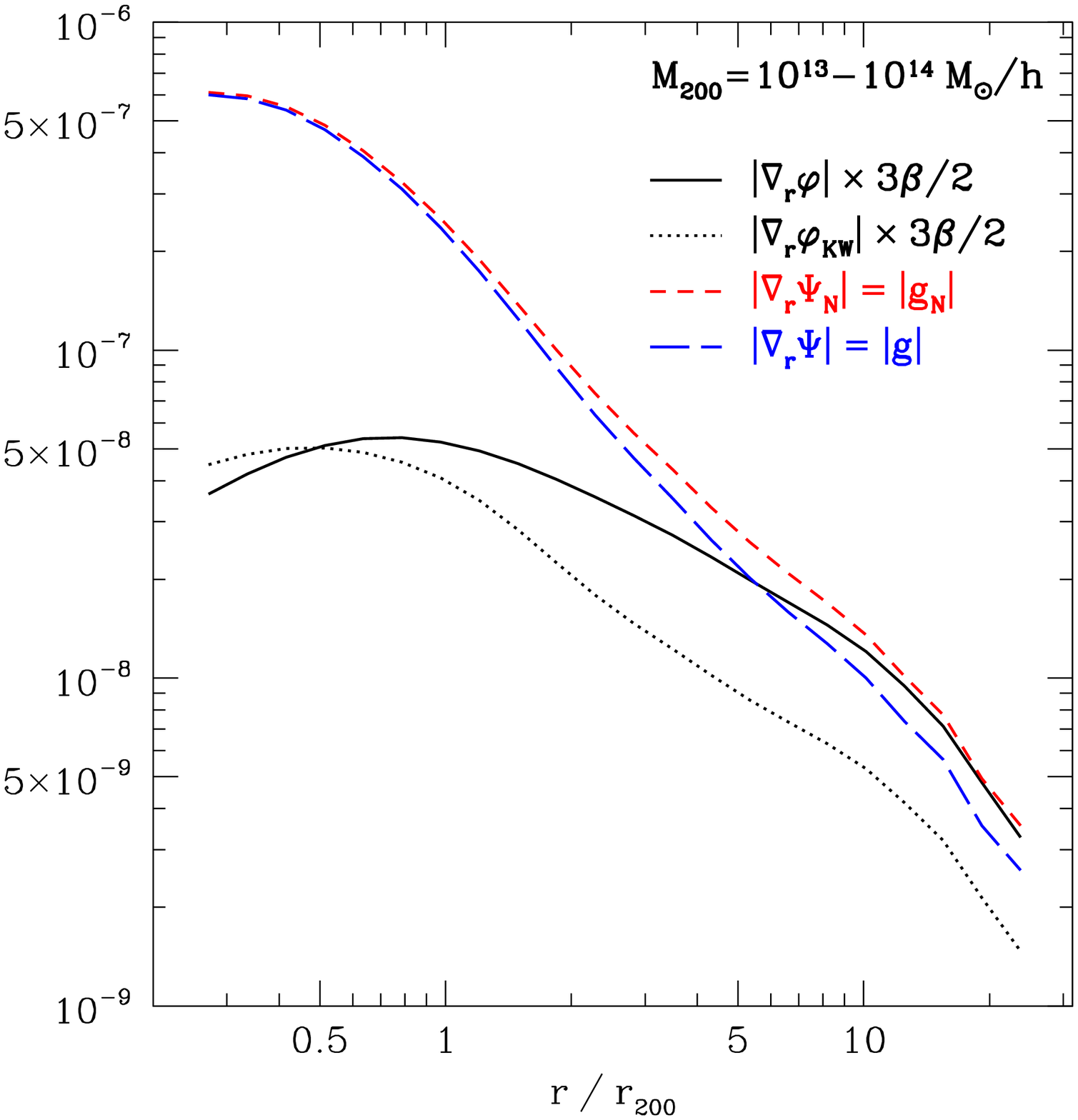}
\caption{{\small \textit{Left panel:} Average halo profiles of $\ph$,
the dynamical potential $\Psi$, and the lensing potential $\Phi_-=\Psi_N$ 
in the DGP simulations at $z=0$, for halos between $10^{13}$ and $10^{14}\Msunh$.
The lower panel shows the average profile of the matter overdensity $\delta$
for the same halos.
\textit{Right panel:} Average halo profiles of the radial gradient of $\ph$,
the DGP potential $\Psi$, and Newtonian potential $\Psi_N$, showing
that $g_r = -\nabla_r\Psi\rightarrow g_N = -\nabla_r\Psi_N$ in the inner regions of halos, i.e. GR is 
restored. Also shown is the brane bending mode calculated
in the approach of \cite{KW}, $\ph_{\rm KW}$ (see text), for the same
halos as in the left panel.
\label{fig:profiles}}}
\end{figure}

\reffig{profiles} (left panel) shows the average profiles of
the brane-bending mode $\ph$, the dynamical potential $\Psi$,
and the lensing potential $\Phi_-$ of dark matter halos in the
full DGP simulations. The brane-bending mode has been scaled
by $3\beta/2$, so that it agrees with $\Phi_-=\Psi_N$ for
the linearized solution.
We only use our highest resolution simulation
box for this measurement ($\Lbox=64\Mpch$). The inner radius
of the profiles is given by one grid cell. The profiles were
measured by spherically averaging many lines of sight for each halo
with $10^{13}\Msunh < M_{200} < 10^{14}\Msunh$,
and then stacked, scaling each profile to the radius $r_{200}$ of
the respective halo. This was done in order to reduce the 
significant scatter in the profiles.

While the lensing potential $\Phi_-$ obeys the standard Poisson
equation in DGP, the dynamical potential $\Psi$ receives
a contribution from the brane bending mode [\refeq{psi}]. 
Apparently, the $\ph$ field flattens out towards the halo center,
which is in qualitative agreement with the non-linear field
solution for a spherical mass (see, e.g. \reffig{sphericalmass}).
Note that $\ph$ turns away from the linear solution when
the overdensity is of order a few, in agreement with
the estimate in \refsec{philin}.

The quantity which is actually observable however is the gradient
of the field, shown in \reffig{profiles} (right panel), which
gives essentially the deviation of the acceleration from the Newtonian
acceleration. For each halo, we measure
the radial gradient relative to the halo center. We again scaled
$\nabla_r\ph$ by $3\beta/2$ to match the linearized solution
to $\nabla_r\Psi_N$.
From the halo exterior towards the interior, the gradient of $\ph$ 
grows initially, turns around at $r \sim 0.8\,r_{200}$, and shrinks for even
smaller radii. As is clearly shown by comparing the dynamical potential
$\Psi$ with its Newtonian value, $\Psi_N$, any observable deviations from GR
are indeed suppressed within massive halos. We find that for
halos between $10^{12}$ and $10^{13}\Msunh$, the deviation in the
acceleration is around $\Delta g/g_N \approx 0.2/(3\beta)$ in the
inner parts, while
it is suppressed down to $\Delta g/g_N \approx 0.05/(3\beta)$ for
halos around $10^{14}\Msunh$. This trend with mass is expected,
given that $r_*\propto M^{1/3}$ for the spherically symmetric solution.

In order to model the behavior of the brane bending mode in the cosmological
context, we can make use of the two limiting cases presented
in \refsec{DGP}. One is the 
spherically symmetric solution (\refsec{vainshtein}), which can
be used in modeling a spherical collapse of dark
matter halos in DGP \cite{LueEtal04}. Moreover, Khoury and Wyman \cite{KW}
have based the approximate field solution in their N-body simulation
on this limiting case.
More realistically however, cosmic structure forms by collapsing into
non-radially symmetric structures such as filaments. The plane wave
density perturbation can serve as another, albeit rather extreme, test case.
For such a perturbation, the non-linear self-coupling of $\ph$ vanishes
(\refsec{planewave}). Hence, we expect the Vainshtein effect
to be typically weakened in a non-radially symmetric setting, compared
to the spherically symmetric case.

In their ansatz, Khoury and Wyman assumed that the spherically symmetric
solution \refeqs{g_sph}{Delta} holds wherever the field becomes non-linear.
Since in this solution, the gravitational constant is rescaled locally
by $1+\Delta(r)$, the Poisson equation for $\Psi$ can be written 
in this ansatz as:
\be
\nabla^2\Psi = 4\pi G\,[ 1 + \Delta(\delta) ]\,a^2\bar\rho_m \delta,
\label{eq:PsiKW}
\ee
where $\Delta$ is now a function of the local overdensity.
Substituting $\delta \sim M/\bar\rho_mr^3$ in \refeq{rstar} [or \refeq{NLest}]
we see again that $(r_*/r)^3 \sim (H_0r_c)^2\beta^{-2}\Om\delta$, with
an order unity coefficient. Khoury and Wyman \cite{KW} chose:
\be
\left(\frac{r_*}{r}\right)^3 = \eps = \frac{8 H_0^2r_c^2}{9\beta^2}\Om\delta
\quad\Rightarrow\quad
\Delta(\delta) = \frac{2}{3\beta}\left(\sqrt{1+\eps}-1\right)/\eps.
\label{eq:DeltaKW}
\ee
Hence, in the DGP model we simulated, $\eps\approx 0.3\,\delta$.
Given the density field in our DGP simulations, we can solve
\refeq{PsiKW} by Fourier transform as done in \cite{KW}, subtract
the Newtonian potential, and compare the brane bending mode $\ph_{\rm KW}$
from this ansatz
with our numerical solution of the full $\ph$ equation. 

\reffig{profiles}
(right panel) shows the result for stacked halo profiles of 
the radial gradient of $\ph_{\rm KW}$.
While $\ph_{\rm KW}$ shows a qualitatively similar behavior to
the full solution,
the suppression due to the Vainshtein effect appears to set in 
at significantly larger radii in $\ph_{\rm KW}$ than in the full solution.
This is qualitatively in agreement with our expectation that 
non-radially symmetric configurations experience a weaker 
non-linear suppression. The fact that $\ph_{\rm KW}$ does not
approach the linear solution at large radii is a result of
the approximation \refeq{PsiKW}, as was derived in the appendix of \cite{KW}.
The precise tensorial structure of \refeq{phiQS} restores $\ph$ to
the linear solution at large distances $r\gg r_V$.
It would be worth investigating whether this simplified (and computationally
much less expensive) approach can be extended to recover the
linear $\ph$ solution at large scales, an essential feature
of the full brane-bending mode interactions.



Although we did not compare our results with a full simulation based
on the spherically symmetric approach of \cite{KW}, these results
seem to indicate that this approximation overestimates the non-linear
suppression of the brane-bending mode, which might affect observables
such as the power spectrum or halo mass function.
We point out however that the crossover scale $r_c$ in the braneworld-inspired model 
considered in \cite{KW} is an order of magnitude smaller than our $r_c$,
and the non-linearity in \refeqs{PsiKW}{DeltaKW} only becomes effective
at much higher overdensities $\delta = O(100)$, so that for the models
studied in \cite{KW}, the solution
is possibly less sensitive to this approximation.

\section{Constraints on the self-accelerating DGP model}
\label{sec:constraints}

We now briefly discuss the impact of our simulation results on
cosmological constraints on the self-accelerating DGP model without $\Lambda$.
First, regarding, \textit{baryon acoustic oscillations}, our results for the 
DGP power
spectrum on quasi-linear scales show that non-linear effects are not
significantly enhanced at the BAO scale. We estimate that any
modified gravity corrections to the BAO scale are below the percent level,
and hence well within the uncertainties of current BAO measurements 
\cite{EisensteinEtal05,PercivalEtal07}.
Including baryon acoustic oscillations will increase the power of the combined CMB and Supernova
constraints on the self-accelerating DGP model \cite{FangEtal} with non-zero
curvature to $4-5\sigma$ \cite{WaynePrivate}.

Second, \textit{weak lensing} measurements constrain the amplitude of the 
matter power spectrum today, which in this model is 
significantly smaller due to the suppression of growth by the earlier
onset of acceleration and the repulsive force mediated
by the brane-bending mode. The {\it linear} power spectrum normalization
at $z=0$ in the models we simulated is $\sigma_8(\rm QCDM)=0.56$ and 
$\sigma_8(\rm DGP)=0.52$, while for a $\Lambda$CDM model with the same 
primordial normalization, it is $\sigma_8(\Lambda\rm CDM)=0.66$.
We found that the non-linear matter power spectrum in the full DGP simulations
is always below that of QCDM (up to $k \lesssim 6\iMpch$, \reffig{PkDGP}).
Hence, the suppression of the non-linear power spectrum in the 
self-accelerating DGP model corresponds to a reduction in the inferred
linear normalization $\sigma_8$ today of at least 0.1, with respect
to a $\Lambda$CDM model with the same initial conditions.
Such a deviation is disfavored by about 1.5 standard deviations by current 
weak lensing measurements \cite{FuEtal}.

The abundance of massive dark matter halos is probed by 
\textit{cluster surveys}. We showed in \refsec{massfct} that the
abundance of halos above $10^{14}\Msunh$ is suppressed by around 20\%
relative to QCDM, which itself only predicts half as many halos in that
mass range as a $\Lambda$CDM model with the same primordial power. The latter
again corresponds to a shift in $\sigma_8$ by 0.1, which is 
constrained to more than $4\sigma$ by X-ray cluster measurements
\cite{Vikhlinin}, when taking into account the systematic errors.

\section{Conclusions}
\label{sec:concl}

The N-body simulations of DGP gravity we presented here show how
the self-interactions of the brane bending mode $\ph$, which are responsible
for restoring General Relativity in dense environments, influence
the formation of structure in the universe. 
In the self-accelerating
DGP model we simulated, this scalar field mediates a repulsive force,
effectively weakening gravity. We indeed find that the field solution turns
away from the linearized solution whenever the matter overdensity $\delta$
becomes of order a few, and that the repulsive force is suppressed by
more than an order of magnitude compared
to its linearized value in the center of massive halos. 
We also compared our full solution to that
obtained in the approximate ansatz of \cite{KW}. While their ansatz
agrees well with our results in the densest regions, it seems to
overestimate the non-linear suppression in less dense regions, such as
the outer regions and environments of halos. This is in line with
our finding that the non-linear suppression of $\ph$ is weaker for 
non-spherically symmetric configurations.

The non-linear matter power spectrum in the DGP simulations
shows that the suppression due to the repulsive $\ph$ field is amplified
on non-linear scales for the linearized DGP simulations. 
In the full DGP simulations, this suppression
is smaller, and eventually turns around on Mpc scales.
The deviations between linerized and full DGP power spectra are noticeable
already on quasi-linear scales, $k \gtrsim 0.1\iMpch$.
At the BAO scale, modified gravity effects on the power spectrum are
at the percent-level.

We found that the abundance of massive dark matter halos is significantly
suppressed in the DGP simulations, compared to a standard gravity
simulation with the same expansion history. Again, the non-linear
suppression of the $\ph$ field alleviates this suppression, and hence
has to be taken into account when using observations to constrain
this type of modified gravity. The effect on the halo mass function
is in fact large enough to make this an interesting observational
probe of more general braneworld scenarios. 

Independently of the
CMB constraints \cite{FangEtal}, our results on the non-linear structure
formation strongly constrain the self-accelerating DGP model 
(without $\Lambda$).
In the future, we plan to extend our simulations to the normal branch
of DGP, and more general braneworld(-inspired) models \cite{deRham,KW,galileon}. The
key part of the simulations, solving for the non-linear interactions
of the brane-bending mode, is generic to all of these models, and
hence the code will be readily generalized to these cases.
We also plan to attempt a modeling of the modified 
gravity effects, in the context of the halo model for example.
This model can then
serve as a framework for cosmological constraints on braneworld gravity
which take into account the non-linear mechanisms inherent to this
model consistently. Simulations of both $f(R)$ gravity and DGP have shown that
most interesting phenomena appear on scales of 1 to tens of Mpc, and in the
abundance and environments of clusters. Fortunately, these scales 
are well accessible
to observations, e.g., through weak lensing, the Lyman-$\alpha$ forest,
and cluster surveys, which can then be used as precision probes of gravity
on cosmological scales.

\acknowledgments

I am indebted to Scott Dodelson, Wayne Hu, and Andrey Kravtsov
for invaluable input and guidance. I would like to thank Hiro Oyaizu
and Marcos Lima for our previous collaborative work on the code and analysis,
and Angela~Olinto, Mike Gladders, Cora Dvorkin, Sam Leitner, Michael Mortonson,
and Amol Upadhye for discussions. The support of the Fermilab computing
staff is gratefully acknowledged.

The simulations used in this work have been performed on the Joint 
Fermilab - KICP Supercomputing Cluster, supported by grants from Fermilab,
Kavli Institute for Cosmological Physics, and the University of Chicago. 
This work was supported by the Kavli Institute for Cosmological 
Physics at the University of Chicago through grants NSF PHY-0114422 and 
NSF PHY-0551142. 

\appendix

\section{Code implementation and discretization}
\label{app:code}

This appendix describes details of the N-body code implementation, focusing
on the relaxation solver for \refeq{phiQS} as the main non-standard
part of the code. The code is written in C++, uses OpenMP
for parallelization of most time-critical operations, and employs FFTW
\cite{fftw} for the Fast Fourier Transforms.

\subsection{Code units}

The comoving code units follow the convention used in 
\cite{shandarin80a,kravtsov97a}, where
\be
\tilde{\v{x}} = a^{-1} \frac{\v{x}}{r_0},\ \ \tilde{t} = \frac{t}{t_0},\ \ 
\tilde{\v{p}} = a \frac{\v{v}}{v_0},\ \ 
\tilde{\Psi} = \frac{\Psi}{\psi_0},\ \ \tilde{\rho} = a^3 \frac{\rho}{\rho_{m0}},
\ee
and
\be
r_{0} = \frac{L_{\rm box}}{N_g},\ \ t_{0} = H_{0}^{-1},\ \ 
v_{0} = \frac{r_{0}}{t_{0}},\ \ \rho_{m0} = \Om\, \rho_{\rm crit, 0},\ \ 
\psi_{0} = v_{0}^{2}. 
\ee
In the above definitions, $L_{\rm box}$ is the comoving simulation box size 
in $\Mpch$, $N_g$ is the number of grid cells in each direction, $H_{0}$ is 
the Hubble parameter today, $\rho_{\rm crit,0}$ is the critical density today, 
and $\Om$ is the fraction of non-relativistic matter today relative to the 
critical density.

Transformed to code units, the equation for $\ph$ becomes:
\be
L(\ph) \equiv \nablat^2 \ph + \frac{\tilde{r_c}^2}{3\beta\,a^2} [
(\nablat^2\ph)^2 - (\nablat_i\nablat_j\ph)^2 ] = f
\equiv \frac{\Om}{a\,\tilde{c}^2\,\beta} \tilde{\delta}.
\label{eq:phicode}
\ee
Here, $\nablat$ acts with respect to code coordinates, $\tilde{r_c} = r_c/r_0$,
$\tilde{c}=1/v_0$, and $\tilde\delta = \delta\rho/\overline{\rho_m}$ is the 
matter overdensity on the grid, determined from the particle positions. 
Note that we have not yet scaled $\ph$ to $\psi_0$.
After solving \refeq{phicode}, $\ph$ is added to the standard Newtonian
potential, which is solved for using a Fast Fourier Transform, to obtain
the dynamical potential in DGP:
\be
\tilde\Psi = \tilde\Psi_N + \frac{1}{2}\frac{\ph}{\psi_0}.
\ee

\subsection{Discretization of $\ph$ equation}

We employ standard second-order symmetric differences for the discretization
of the second derivatives in \refeq{phicode}:
\bea
\nablat_x\nablat_x\ph_{i,j,k} &=& \frac{1}{h^2} \left (
\ph_{i+1,j,k} + \ph_{i-1,j,k} - 2\ph_{i,j,k} \right ), \label{eq:disc1}\\
\nablat_x\nablat_y\ph_{i,j,k} &=& \frac{1}{4h^2} \left (
\ph_{i+1,j+1,k} - \ph_{i+1,j-1,k} - \ph_{i-1,j+1,k} + \ph_{i-1,j-1,k} \right ),
\label{eq:disc2}
\eea
and correspondingly for derivatives with respect to $y$, $z$. Here, $i,j,k$
stand for the grid indices, and the step size $h=1$ for the base grid and 
$h=2^n$ for the grid of refinement level $n$. The finite differences are
evaluated before squaring in calculating the non-linear terms in \refeq{phicode}.

We have tried different discretizations, such as going to higher order
in the finite differences, and solving for the {\it deviation} of $\ph$
from the solution of the linearized equation [\refeq{philin}], instead
of solving for $\ph$ itself. The 
performance of those discretizations is comparable to or worse than
the simple discretization \refeqs{disc1}{disc2}. This is understandable, since
going to higher order amounts to making the derivative operations less local
in order to be sensitive to error modes of longer wavelength. 
However, in our multigrid relaxation scheme, this is already done efficiently by
the coarser grids, so going to higher order in a single relaxation does
not improve performance. Hence, we stay with the straightforward
discretization, \refeqs{disc1}{disc2}.

\subsection{Relaxation algorithm}

In general, a relaxation scheme operates by iteratively obtaining
a better approximation to the solution $\ph^{(i+1)}$ given a previous
guess, $\ph^{(i)}$. On the grid, the solution is updated successively
for each cell $(i,j,k)$:
\be
\ph^{(i+1)}_{i,j,k} \leftarrow F(\ph^{(i)}_{l,m,n}),
\label{eq:relax}
\ee
where $F$ is the discretized field equation solved for $\ph_{i,j,k}$.
In case of a linear field equation, $F$ can be solved for $\ph_{i,j,k}$
straightforwardly.
However, in our case the field equation is non-linear in $\ph$, and
we instead solve for $F$ iteratively via Newton's method. Writing
the field equation \refeq{phicode} as $L(\ph) = f$, we need to solve:
\be
L(\dots,\,\ph_{i,j,k},\,\dots) - f_{i,j,k} = 0
\ee 
for $\ph_{i,j,k}$, where we have suppressed the dependences of $L$ on 
neighboring grid cells. By expanding $L(\ph_{i,j,k})$ in a Taylor series
around the current approximation, one step of Newton's method works
by updating $\ph^{(i)}_{i,j,k}$ with:
\be
\ph^{(i+1)}_{i,j,k} \leftarrow \ph^{(i)}_{i,j,k} - \frac{L(\dots,\,\ph^{(i)}_{i,j,k},\,\dots) - f_{i,j,k}}{
\partial L/\partial \ph^{(i)}_{i,j,k}}.
\label{eq:relaxNewton}
\ee
Our relaxation step is thus given by \refeq{relaxNewton}, where
$\partial L/\partial \ph_{i,j,k}$ is determined from \refeq{phicode}
and the discretization \refeqs{disc1}{disc2}.
In practice, the order in which we loop over grid cells in performing
the relaxation \refeq{relaxNewton} is important, since each relaxation
step depends on the neighboring grid cells. In particular, we need to
take care when parallelizing the relaxation. We use a generalized red-black
scheme (see, e.g., \cite{Briggs,HPMpaper}), by successively running over
cells with $(i+j+k)$~modulo~4 = $n$, $n=0,1,2,3$. This was done in order
to break dependences due to neighboring cells within each set, in particular
due to the mixed derivatives \refeq{disc2}, so that each set can 
be efficiently parallelized. We experimented with different ordering
schemes and found that this choice performed best.

Now, assume we have an approximate solution $\ph$ to the field equation
\refeq{phicode}
on the fine grid $\Omega^h$, which differs from the true solution $\bar\ph$
by the {\it error} $e$, $\bar\ph = \ph - e$.
Further, the {\it residual} is defined as $r = L(\ph)-f$. Since $L(\bar\ph)=f$
by assumption, we obtain the following \textit{residual equation}:
\be
L(\ph) - L(\ph - e) = r.
\label{eq:res}
\ee
We expect that the relaxation on the fine grid has removed
most small-scale error modes, so $e$ mainly consists of longer wavelength
modes. Thus, we will solve for $e$ on the coarser grid $\Omega^{2h}$,
and then correct the approximate solution $\ph$ for this error. On the coarse 
grid $\Omega^{2h}$, \refeq{res} reads:
\be
L^{2h}(I^{2h}_h\ph^h) - L^{2h}(I^{2h}_h\ph^h - e^{2h}) = I^{2h}_h 
\left ( L^h(\ph^h) - f^h \right ).
\label{eq:res2h}
\ee
Here, the superscripts denote which grid a given quantity is defined on,
and $I^{2h}_h$ is the restriction operator mapping fields from the fine
grid to the coarse grid. Note that \refeq{res2h} is of the same form
as the $\ph$ equation on the coarse grid, $L^{2h}(\ph^{2h})=f^{2h}$, except
with a different right-hand side. Hence, \refeq{res2h} is solved for $e^{2h}$
using the same algorithm as the original equation, but at one grid level higher.
Once \refeq{res2h} is solved, we can correct $\ph^h$ for the error:
\be
\ph^h \leftarrow \ph^h - I^h_{2h}e^{2h},
\label{eq:correct}
\ee
where $I^h_{2h}$ is the interpolation operator mapping the error from the
coarse grid to the fine grid.

In summary, the multigrid relaxation proceeds as follows:
\begin{itemize}
\item Relax \refeq{phicode} on the fine grid $\Omega^h$ $N_I$ times. Compute the residual
$r^h = L^h(\ph^h)-f^h$.
\item Solve \refeq{res2h} for the error $e^{2h}$:
\begin{itemize}
\item Relax \refeq{res2h} on $\Omega^{2h}$ $N_I$ times. Compute the residual
(l.h.s.--r.h.s.) of \refeq{res2h}.
\item Solve \refeq{res2h} for $e^{4h}$:
\begin{itemize}
\item[...]
\end{itemize}
\item Correct $e^{2h}$ using $I^{2h}_{4h}e^{4h}$ [\refeq{correct}].
\end{itemize}
\item Correct $\ph^h$ using $I^{h}_{2h}e^{2h}$ [\refeq{correct}].
\end{itemize}
Thus, the relaxation proceeds as a nested loop going down to the coarsest
grid, and then correcting the solutions on the successively finer grids.
One such iteration through all grid levels is called a \textit{V-cycle}.
The coarsest grid we use has 4 cells on a side, which corresponds to
a refinement of $2^7$ for a $N_g=512$ base grid.

For the interpolation operator $\Omega^{2h}\rightarrow\Omega^h$, we use 
standard bilinear interpolation (consistent with the cloud-in-cell scheme
used for assigning densities and measuring accelerations on the grid).
For the restriction $\Omega^h\rightarrow\Omega^{2h}$, we use full-weighting,
the transpose of the bilinear interpolation operator. See the appendix
in \cite{HPMpaper} for an explicit definition.

In order to ensure convergence at all times, we adopt a large value
of $N_I=10$, at the expense of computing time. Usually, one V-cycle
reduces the residual $\L = \sqrt{\langle r^2 \rangle}$ by 1--2 orders
of magnitude. We stop the relaxation when either $\L < 10^{-14}$ is reached,
or the convergence stalls at a certain level of $\L$. Either of this
usually happens within 4 V-cycles. The value of $\L$ for each time 
step is logged, allowing for a monitoring of the convergence status in
the simulations.

\section{Sine wave test}
\label{app:sinetest}

\begin{figure}[t!]
\centering
\includegraphics[width=0.48\textwidth]{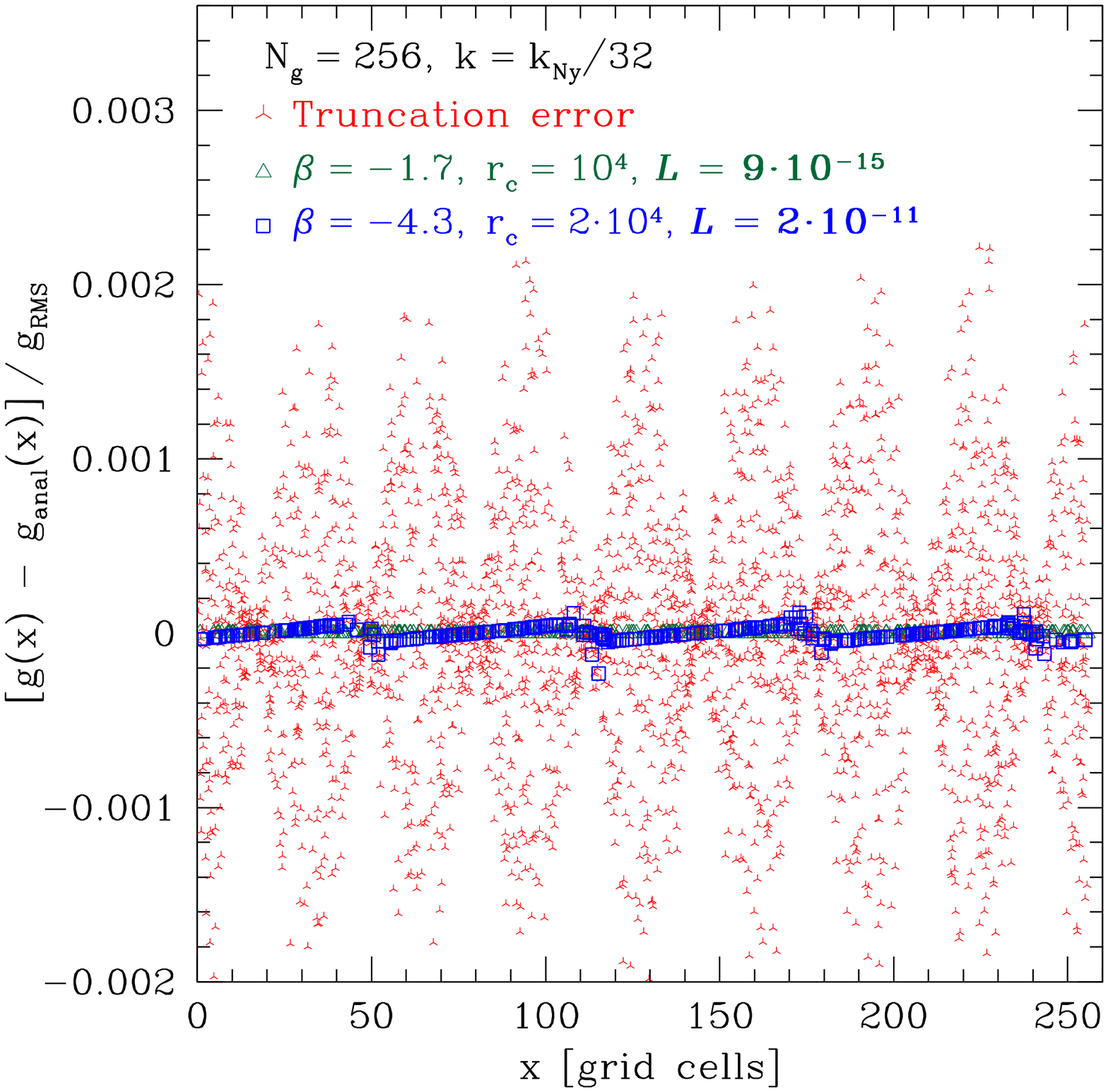}
\includegraphics[width=0.48\textwidth]{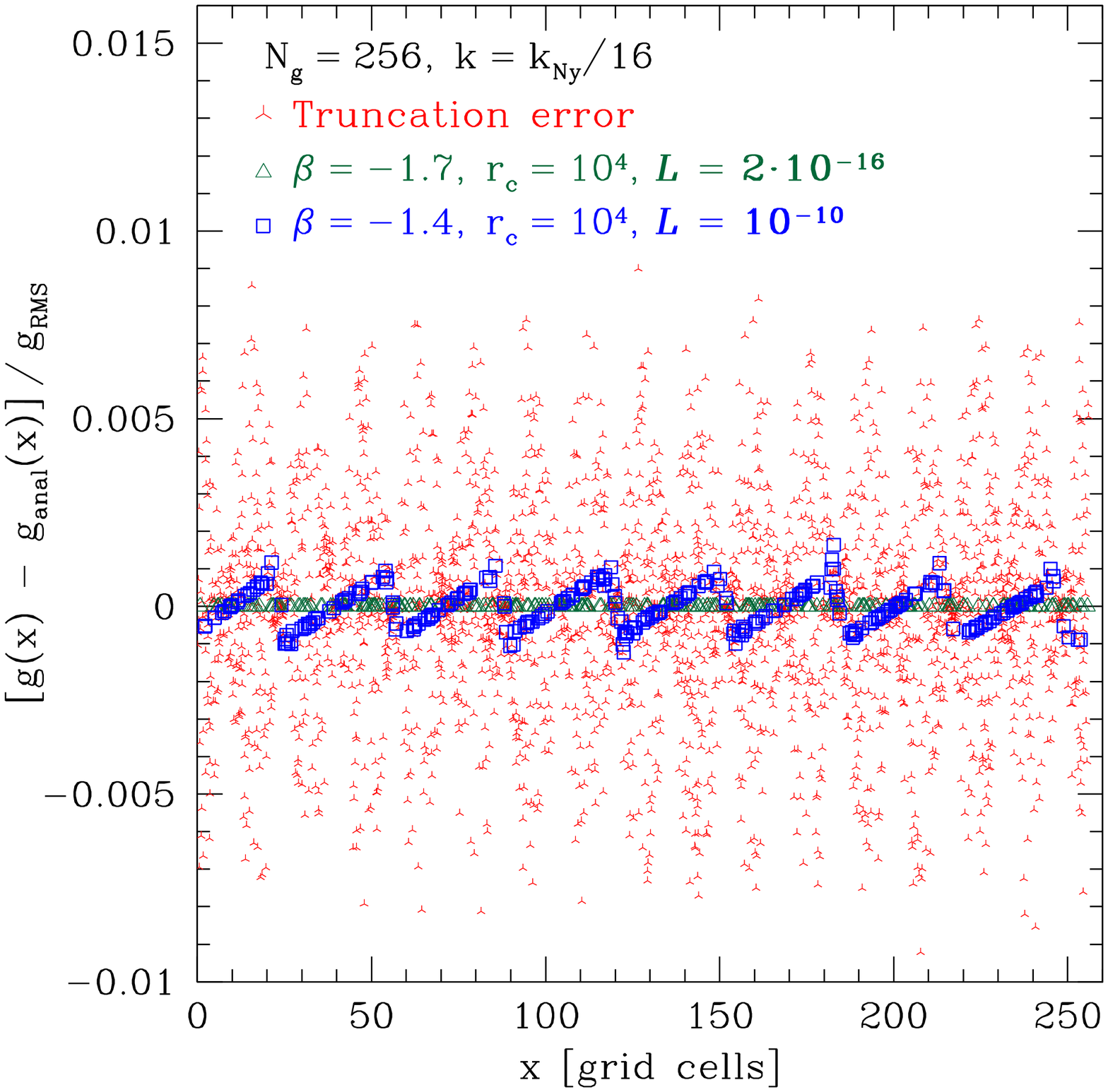}
\caption{{\small Truncation errors (red points) and residual errors 
in the $\ph$ field solution (green, blue points) for sine waves with
$k=\kNy/32$ \textit{(left panel)} and $k=\kNy/16$ \textit{(right panel)}.
The parameters used for the field equation are shown
(with $r_c$ in $\Mpch$), as well as the RMS residuals $\L$ of the final
solution. The errors are shown as deviations of the acceleration
measured at random points from the expectation, scaled to
the RMS acceleration of the sine wave (see text).}
\label{fig:sinetest}}
\end{figure}

In this section, we use the fact that the full $\ph$ solution is
identical to the linearized solution for a plane wave perturbation
(\refsec{planewave})
to test our code for perturbations on different length scales.
We consider a simple sine wave density perturbation, with $k$-vector chosen
to lie along the $x$-axis:
\be
\delta(x) = \frac{\delta\rho}{\rho} = A\,\sin k\, x.
\ee
In this case, the solution to the non-linear $\ph$ equation is identical to the 
linear solution:
\be
\ph(x) = \frac{2}{3\beta} \Psi_N(x) = \frac{2}{3\beta}\frac{3\,\Om}{2\,a\,k^2}
A\,\sin k\,x.
\label{eq:phisine}
\ee
Hence, the exact acceleration is given by:
\be
g(x) = \left ( 1 + \frac{1}{3\beta} \right ) g_N(x) = 
\left ( 1 + \frac{1}{3\beta} \right )\frac{3\,\Om}{2\,a\,k}A\,\cos k\,x.
\label{eq:gsine}
\ee
For a given level of residuals $\L$, we now demand that the errors due
to the approximate solution of the $\ph$ equation of motion are small
compared to the unavoidable truncation errors which are encurred by
taking the gradient of the potential on the grid.

In the following, we set $a=1$, $\Om=0.3$, and the number of grid cells
is set to $256^3$.
The truncation errors are measured by assigning the analytical solution
for the potential to each grid point, and comparing the exact acceleration
to the one obtained by derivation and interpolation from the grid. The
truncation errors for two sine waves of different wavelength are shown
as deviations in the acceleration in \reffig{sinetest} (red points). 
We scale the deviations to the RMS acceleration of the sine wave,
$g_{\rm RMS} = 3\Om/(2\sqrt{2}\,a\,k)\,A$, in order to avoid far outliers
obtained near the zeros of the sine wave when scaling to the exact solution
at each point. For shorter wavelength modes,
the truncation error increases and reaches order unity for waves on
the Nyquist scale, as expected.

In order to estimate the additional error in the acceleration made
due to insufficient convergence of the solution, we compare the 
acceleration measured from the non-linear field solution coming out of
the relaxation solver, to the corresponding solution 
of the linearized $\ph$ equation \refeq{philin}. We vary the parameters
$\beta$ and $r_c$ which control the strength of the non-linearity in
\refeq{phiQS}. In \reffig{sinetest}, we show the measured
deviations in the acceleration from the linear solution for different
values of $\beta$ and $r_c$, and in particular for the largest values
that reached convergence. The final residuals of the $\ph$ solution in
each case are also shown.

Apparently, the errors in the acceleration are completely negligible
for residuals $\L \leq 10^{-14}$, consistent with results we found for
other test cases such as the spherical mass. For sufficiently strong
non-linearity and short wavelengths, residuals of order 
$\L\sim 10^{-11}-10^{-10}$ are reached which lead to measurable
deviations in the acceleration. However, for the residual values up to 
$\sim 10^{-10}$ that we explored, the error due to the incomplete convergence
is very small compared to the truncation error in all cases. Hence,
we conclude that for residuals smaller than the conservative bound of
$10^{-10}$, the $\ph$ solution we
obtain is sufficiently accurate in order not to bias the particle dynamics.

\begin{figure}[t!]
\centering
\includegraphics[width=0.48\textwidth]{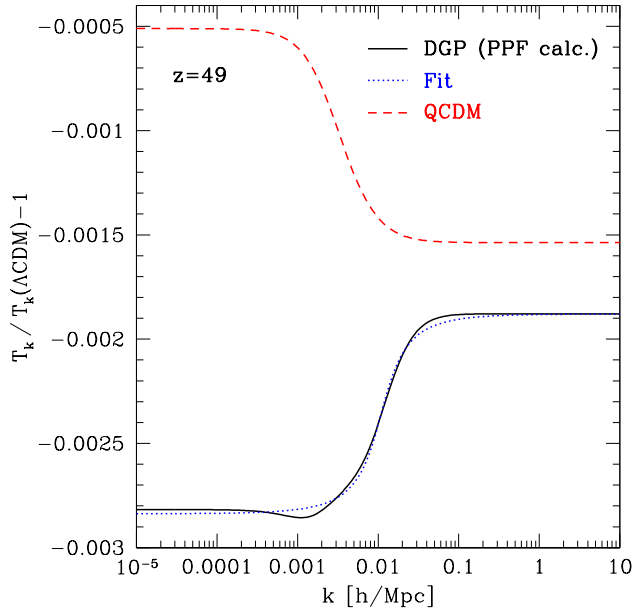}
\caption{{\small Deviation of the DGP and QCDM matter transfer functions from
$\Lambda$CDM at the initial redshift of the simulations, $z_i=49$.
The solid line shows the PPF calculation for DGP, while the blue dotted
line is a simple fit used to correct the initial conditions (see text).}
\label{fig:TkDGP}}
\end{figure}
\section{Correcting Initial Conditions for DGP}
\label{app:ICcorr}

Due to the additional term $H/r_c$ in the Friedmann equation, the
self-accelerating DGP model starts to deviate from a $\Lambda$CDM expansion
history at relatively high redshifts. In addition, there are modifications
to the growth of horizon-scale modes. For this reason, even at
the initial redshift of our simulations, $z_i=49$, there are small
departures in the matter transfer function in DGP as compared to a 
$\Lambda$CDM model with the same early-Universe parameters 
$\Om=1-\OL,\Ob,h,n_s,A_s$.

In order to take these differences into account, we calculate the dark matter
transfer function $T(k)$ at $z_i$ for DGP and QCDM, using the PPF approach described in
\cite{hu07b,FangEtal}, and for the corresponding flat $\Lambda$CDM
model. We do not include the effects of radiation on the transfer function,
which is not necessary since we are only interested in 
the relative deviation of the DGP $T(k)$ from $\Lambda$CDM.
\reffig{TkDGP} shows the relative deviation in $T(k,z_i)$ from $\Lambda$CDM for DGP and QCDM.
Both QCDM and DGP transfer functions are slightly suppressed on all scales
due to the effects of the earlier onset of acceleration.
On super-horizon scales, the transfer function is further suppressed in DGP
caused by the transition to 5D gravity on very large scales. 
In contrast, $T(k)$ is less suppressed in 
QCDM because of the fluctuations in the effective dark energy.

We correct the transfer function obtained from CAMB for these small 
differences at $z_i$ by multiplying $T_{\rm CAMB}(k)$ with $1+f(k)$,
where $f(k)$ is a simple $\arctan+\rm const.$ function fit to the
DGP curve shown in \reffig{TkDGP}. Hence, these initial conditions
are not quite correct for the QCDM simulations. However, the differences
are very small on the scales probed by our simulations ($k\gtrsim 0.01\iMpch$).

\bibliography{DGPM}

\end{document}